\documentclass[twocolumn,tighten,times,floatfix]{aastex631}
\usepackage{amsmath}
\usepackage{xspace,xcolor}
\usepackage{natbib}

\newcommand{\beq}{\begin{equation}}
\newcommand{\eeq}{\end{equation}}
\newcommand{\igm}{\texttt{IllinoisGRMHD}\xspace}
\newcommand{\etk}{\texttt{Einstein Toolkit}\xspace}
\newcommand{\carpet}{\texttt{Carpet}\xspace}
\newcommand{\newrad}{\texttt{Newrad}\xspace}
\newcommand{\baikal}{\texttt{Baikal}\xspace}
\newcommand{\harmnuc}{\texttt{HARM3D+NUC}\xspace}

\newcommand{\lorene}{\texttt{LORENE}\xspace}
\newcommand{\Fdual}{{}^{*}\!F}
\newcommand{\rhob}{\ensuremath{\rho_{\rm b}}\xspace}
\newcommand{\mb}{\ensuremath{m_{\rm b}}\xspace}
\newcommand{\nb}{\ensuremath{n_{\rm b}}\xspace}
\renewcommand{\ne}{\ensuremath{n_{\rm e}}\xspace}

\newcommand{\ye}{\ensuremath{Y_{\rm e}}\xspace}
\newcommand{\Msun}{\ensuremath{M_{\odot}}\xspace}

\newcommand{\ISEF}{ISEF International Fellowship}
\newcommand{\NASAFS}{NASA Einstein fellow}
\def\eqref#1{Eq.\;(\ref{#1})\xspace}
\newcommand{\sfho}{\texttt{SFHo}\xspace}

\begin{document}

\author[0000-0002-0632-8897]{Yossef Zenati}
\altaffiliation{\ISEF}
\affil{Physics and Astronomy Department, Johns Hopkins University, Baltimore, MD 21218, USA}

\author[0000-0002-2995-7717]{Julian~H.~Krolik}
\affil{Physics and Astronomy Department, Johns Hopkins University, Baltimore, MD 21218, USA}

\author[0000-0002-4541-8553]{Leonardo~R.~Werneck}
\affil{Department of Physics, University of Idaho, Moscow, ID 83843, USA}

\author[0000-0003-2333-6116]{Ariadna~Murguia-Berthier}
\altaffiliation{\NASAFS}
\affil{Center for Interdisciplinary Exploration and Research in Astrophysics (CIERA), 1800 Sherman Ave., Evanston, IL 60201, USA}

\author[0000-0002-6838-9185]{Zachariah~B.~Etienne}
\affil{Department of Physics, University of Idaho, Moscow, ID 83843, USA}
\affil{Department of Physics and Astronomy, West Virginia University, Morgantown, WV 26506}
\affil{Center for Gravitational Waves and Cosmology, West Virginia University, Chestnut Ridge Research Building, Morgantown, WV 26505}

\author[0000-0003-3547-8306]{Scott~C.~Noble}
\affil{Gravitational Astrophysics Lab, NASA Goddard Space Flight Center, Greenbelt, MD 20771, USA}

\author[0000-0002-7964-5420]{Tsvi~Piran}
\affil{Racah Institute of Physics, Hebrew University, Jerusalem, 91904, Israel}

\title{Bound Debris Expulsion from Neutron Star Merger Remnants}
\correspondingauthor{Yossef Zenati}
\email{yzenati1@jhu.edu}

\date{March 2023}

\begin{abstract}
Many studies have found that neutron star mergers leave a fraction of the stars' mass in bound orbits surrounding the resulting massive neutron star or black hole. This mass is a site of $r-$ process nucleosynthesis and can generate a wind that contributes to a kilonova. However, comparatively little is known about the dynamics determining its mass or initial structure. Here we begin to investigate these questions, starting with the origin of the disk mass. Using tracer particle as well as discretized fluid data from numerical simulations, we identify where in the neutron stars the debris came from, the paths it takes in order to escape from the neutron stars' interiors, and the times and locations at which its orbital properties diverge from those of neighboring fluid elements that end up remaining in the merged neutron star.
\end{abstract}

\keywords{Neutron Star --- General relativity --- Accretion disk --- black hole}

\section{Introduction} \label{sec:intro}

On August 17, 2017, the \texttt{LIGO/Virgo} collaboration (LVC) detected the first gravitational wave (GW) signal from a binary neutron star (BNS) merger, the GW170817 event \citep{Abbott17:gw, villar17}. This was also the first multi-messenger detection of such an event, as it was seen across the electromagnetic (EM) spectrum (Fermi, Chandra, UV, optical, IR, radio).

From GW170817 and similar BNS mergers, we can learn about many fundamental questions in general relativity, cosmology, and astrophysics \citep{Coulter17, Guidorzi17, Haggard17, Margutti+17}.
For example, the distance to this event inferred from GWs was consistent with entirely independent measurements of the Hubble constant \citep{Abbott17a}.
Analysis of the GW170817 GW form led to important constraints on the nuclear equation of state (EOS) \citep{Damour_Nagar_Viilain+12, Radice+18_NSEoS, BernuzziS20}.
Its optical/IR emission (optical/IR analysis papers) gave support to the idea that such mergers could be a main channel of rapid neutron-capture ($r$-process) nucleosynthesis \citep[e.g,][]{Lattimer_Schramm74, Eichler+89, AconesThielemann13, Cowan2021, Pian2023}.
In addition, it also demonstrated that, as predicted many years before, neutron star (NS) mergers can create short $\gamma$-ray bursts \citep[see,][]{Eichler+89, Rosswog_Ramirez-Ruiz02, Nakar07, Berger14, Goldstein+17}

There is a strong theoretical consensus that a portion of the neutron stars' original mass is not held in the remnant, whether it is a massive neutron star or a black hole. Mass can be separated during numerous stages of a BNS merger: the late inspiral phase, the merger stage, and the subsequent evolution of the initially bound debris. Multiple mechanisms can contribute, including tidal gravity, shock heating, neutrino heating, nuclear heating, magneto-centrifugal winds (from both a debris disk and highly-magnetized massive neutron star merger remnant), and magnetohydrodynamic (MHD) turbulent dissipation \citep[see,][]{Metzger+10_KN, Hotokezaka+13a, Fernandez+15, Radice+20_AnnRev, Margutti_Chornock21, Sarin_Lasky21, Most-Quataert2023, Combi-Siegel2023, Shibata+23}. 

Some of the matter that escapes from the remnant is ejected from the system altogether during the late inspiral and the merger itself. 
This part is called ``dynamical ejecta", and it is generally ascribed to a combination of tidal forces and shock heating when the two stars come into direct contact \citep{Hotokezaka+13a}. Its mass is rather small, ${\sim}10^{-3}\Msun$~\citep{Radice+16, SiegelMetzger17PRL, Fujibayashi+18, Shibata+23}.
The dynamical ejecta have received much attention for their possible role in producing a kilonova (KN) event, particularly those features indicating especially high outflow velocities~\citep{Metzger+10_KN, BarnesKasen13, Radice+18, Domoto+21, Curtis+23}.

However, the same studies predict that the rest of the matter left outside the remnant is, at least temporarily, bound; its mass is rather larger, ${\sim}$0.01--0.1\Msun. Here we call this portion the ``bound debris". It, too, likely plays a major role in neutron star merger phenomenology. The magnetic flux and confinement pressure needed by relativistic jets may be supplied by this debris. In addition, on rather longer timescales than dynamical ejection, the heat released inside the bound debris by nucleosynthesis and dissipation of MHD turbulence may drive a wind carrying away neutron-enriched isotopes. The photosphere of such a wind could also contribute to kilonova radiation \citep[e.g,][]{Rosswog_Ramirez-Ruiz02, Fernandez+15, Just+15, Korobkin+20}.


Despite its likely importance to the phenomenology of neutron star mergers, there has been little effort to understand the mechanisms governing how much mass is in bound orbits or the nature of its dynamical state, i.e., its distribution of angular momentum and energy: \citet{Most+21} studied the correlations between the angular momentum distribution of debris and other properties, but only for NS-BH mergers. The distribution of mass with angular momentum has special interest because it determines the initial surface density profile of the bound debris, which is almost certainly far from inflow equilibrium. The initial angular momentum distribution, therefore, governs the bound debris' subsequent evolution.

In this paper, we will begin the study of these issues, starting with the most basic questions: From which part(s) of the NSs is the bound debris drawn?  How, and at what point in the merger, is the bound debris separated from the majority of the merged NS mass, the part that remains in the newly-formed NS or is carried into the black hole (BH) if one is created?

\begin{figure*}[h]
\centering
\includegraphics[width=0.99\linewidth]{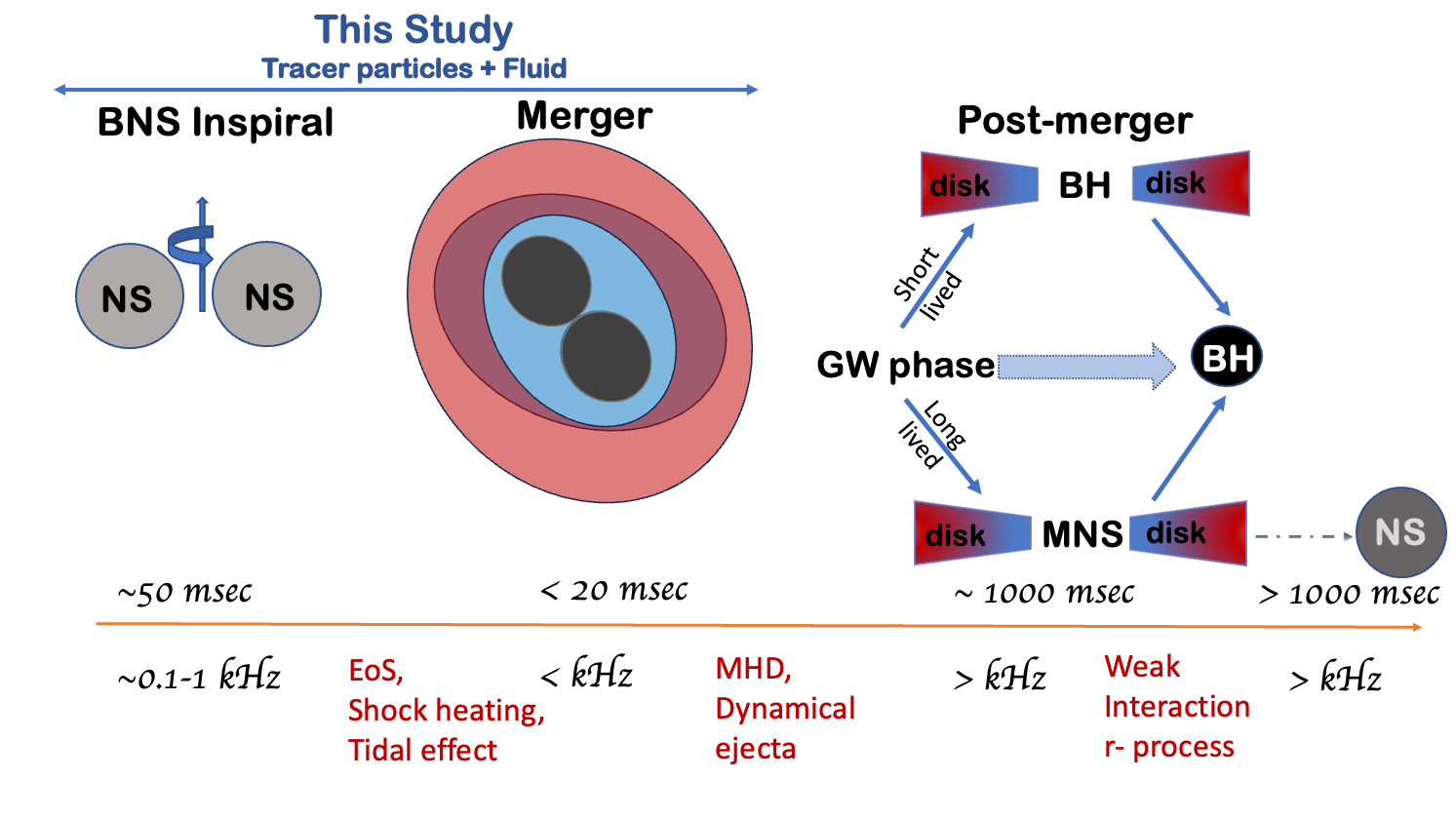}
\caption{Schematic presentation of the different stages and outcomes of BNS mergers: from left to right, inspiral, merger, post-merger remnant plus debris disk; and, possibly, a stable NS. Timescales for each stage are shown below the corresponding sketches. The bottom line lists the GW frequency and the principal physical mechanisms of each stage.}
\label{fig:cartoon}
\end{figure*}

\section{Calculation}
\label{sec:model}

\subsection{Overview}

To place our calculation in context, we first review the stages of a BNS merger (see Fig.~\ref{fig:cartoon}). Once formed, a NS binary can evolve only by emission of gravitational radiation. During very nearly the entire span of this period, the stars' separation is so large that there is no direct interaction between them, and even gravitational tidal forces are negligible. When, however, their separation shrinks to only a few times their radii, tidal forces begin to distort them; this is the period we call ``late inspiral". Eventually, they touch, beginning the ``merger proper", which creates a very massive NS (MNS). Depending upon the mass and rotational state of this newly-formed star, it may collapse to a BH (almost immediately or after some delay) or live for an indefinitely long time. In the course of these events, a small fraction of the initial total mass is expelled.

When the mass distribution is substantially asymmetric (the inspiral, merger, and possible collapse phases), there is no simple description of the spacetime.

Consequently, it is necessary to employ numerical relativity methods in order to follow gravitational dynamics. On the other hand, once the mass distribution relaxes to axisymmetry, either as a stable NS or a Kerr BH, there is no need for a further solution of the Einstein field equations. For this reason, we divide our simulation of such events into two pieces. The first part is performed with the numerical relativity plus MHD code \igm~\citep{Etienne+15, Etienne+20, Werneck+23}; the second part uses \harmnuc~\citep{AriM+21}. A ``hand-off" procedure assures consistency between the two in terms of the spacetime and the fluid properties \citep{Handoff22}. In the present paper, we make use of data taken only from the first part; future work in this project will employ data from the second part for studies of the bound debris' evolution, including jet-launching, nucleosynthesis, and wind-driving.

\subsection{Physics treated and equations solved}

\igm  solves the equations of MHD,
\begin{align}
  \nabla_{\mu}\left(\nb u^{\mu}\right) &= 0\;, \label{eq:baryon_cons}\\
  \nabla_{\mu}\left(\ne u^{\mu}\right) &= {\cal R}\;, \label{eq:lepton_cons} \\
  \nabla_{\mu}T^{\mu\nu} &= \mathcal{Q} u^{\nu}\;, \label{eq:enmom_cons}\\
  \nabla_{\mu}\Fdual^{\mu\nu} &= 0\;, \label{eq:maxwell}
\end{align}
corresponding to the conservation of the baryon number, conservation of Lepton number, conservation of energy-momentum, and the two homogeneous Maxwell's equations, respectively.
In coordination with the solution of these equations, the BSSN evolution thorn \texttt{Baikal} solves the Einstein Field Equations.

In these equations, $\nb$ ($\ne$) is the baryon (lepton) number density and $u^{\mu}$ is the fluid four-velocity. The net rate of change in the lepton number ${\cal R}$ is the difference between the creation rates of electron anti-neutrinos and neutrinos. The cooling rate ${\cal Q}$ is the rate at which neutrinos (and anti-neutrinos) carry energy out of the gas. It is calculated from a local emission model and a leakage rate formalism \citep[see][for details about the calculation of ${\cal R}$ and ${\cal Q}$]{AriM+21, Werneck+23}. 

The energy-momentum tensor is assumed to be that of a perfect fluid plus an EM contribution,
\beq
  T^{\mu\nu} = \left(\rhob h + b^{2}\right)u^{\mu}u^{\nu} + \left(P + \frac{b^{2}}{2}\right)g^{\mu\nu} - b^{\mu}b^{\nu},
\eeq
where \mbox{$\rhob = \mb\nb$} is the baryon density, $\mb$ is the mean baryon mass, \mbox{$\Fdual^{\mu\nu}=(1/2)\tilde{\epsilon}^{\mu\nu\rho\sigma}F_{\rho\sigma}$} is the dual of the Faraday tensor $F^{\mu\nu}$, and $\tilde{\epsilon}^{\mu\nu\rho\sigma}$ is the Levi-Civita tensor.  The parameter
\mbox{$h = 1 + \epsilon + P/\rhob$} is the specific enthalpy, $\epsilon$ is the specific internal energy, and $P$ is the fluid pressure, which is given through a tabulated form of the \sfho EOS for hot degenerate matter~\citep{OConnor:2009iuz}.\footnote{The \sfho EOS table was downloaded from \url{http://stellarcollapse.org}.} 

Further, $b^{\mu}=(4\pi)^{-1/2}B^{\mu}_{(u)}$ is the rescaled magnetic 4-vector in the fluid frame, where
\begin{align}
  B^{0}_{(u)} &= u_{i}B^{i}/\alpha\;,\\
  B^{i}_{(u)} &= \bigl(B^{i}/\alpha + 
               B^{0}_{(u)}u^{i}\bigr)/u^{0}\;.
\end{align}
Here $B^{i}$ is the magnetic field in the frame normal to the hypersurface, $b^2 \equiv b^\mu b_\mu$ is the magnetic energy density, $g_{\mu \nu}$ is the spacetime metric, and $\alpha$ is the lapse function.  The lapse, the shift vector, and the metric are all defined on spatial hypersurfaces of constant coordinate time $t$. Ideal MHD (\mbox{$u_{\mu}F^{\mu}=0$}) is assumed throughout.

In addition to solving these equations, we also follow the locations and velocities of a large number of tracer particles. The initial positions of the tracer particles are selected by randomly choosing points $p=(x,y,z)$ within a radius $R_{\rm seed}=75$~km from the system center of mass and accepting the particle if
\begin{equation}
  \frac{\rhob(x,y,z)}{\max\rhob} > \zeta\;,
\end{equation}
where $\zeta\in[0,1]$ is a random number. The process is repeated iteratively until all $N_{\rm particles}=50,000$ have been seeded.
The fact that the NSs are extremely compact and have central densities many orders of magnitude larger than that of the  surrounding gas ensures that the vast majority of the tracer particles are seeded within each NS, as shown in Fig.~\ref{fig:rho_particles}.


Once the positions \mbox{$p_{n} = (x_{n}, y_{n}, z_{n})$} of the particles are known, where \mbox{$n=0,1,\ldots,N_{\rm particles}-1$}, we update them in time using
\begin{equation}
    \frac{dp^{i}_{n}}{dt} = v^{i}(p_{n})\;,
\end{equation}
where $v^{i}(p_{n})$ is the fluid three-velocity $v^{i} = u^{i}/u^{0}$ interpolated to the particle's position $p_{n}$. The tracer particle positions are updated every 16 local time steps, and the three-velocities are obtained using fourth-order Lagrange interpolation.

\subsection{Numerical setup}

We perform BNS simulations using the \etk~\citep{Loffler:2011ay,roland_haas_2022_7245853} \footnote{See {\url{https://github.com/zachetienne/nrpytutorial}}.} applied to a Cartesian grid with adaptive mesh refinement (AMR) provided by \carpet~\citep{Schnetter:2003rb}. Initially, the grid contains eight refinement levels differing by factors of two, such that the resolution at the finest refinement level is ${\approx}185$\,m. Once the minimum value of the lapse function drops below 0.1, BH formation is assumed to be imminent and two additional refinement levels are added, such that the resolution of the finest refinement level becomes ${\approx}46$\,m.

At the outer boundary, located at ${\approx}5670\ \rm{km}$, we apply radiation boundary conditions to the metric quantities using \newrad~\citep{Alcubierre:2002kk, Loffler:2011ay} and a simple copy boundary condition for the MHD fields.

The spacetime is evolved using \baikal~\citep{roland_haas_2022_7245853}, while the MHD fields are evolved using a newly developed version of \igm~\citep{Werneck+23} that supports finite-temperature, microphysical EOSs, and neutrino physics via a leakage scheme. In addition to the MHD fields, we evolve the entropy by assuming that it is conserved~\citep[see e.g.,][for a similar strategy]{Noble:2008tm}, an approximation that, while poor at shocks, allows us to use the entropy as a backup variable during primitive recovery~\citep[see][for more details]{Werneck+23}.

Initial data for an equal-mass binary is constructed using \lorene~\citep{Gourgoulhon+01, Feo+16,lorene_website}, with each NS having a baryonic (gravitational) mass of 1.550\Msun (1.348\Msun; see Table~\ref{table_mhd_torus}). The initial separation of the stars is 45\,km and the stellar radii are ${\approx}9.3$\,km. Each is given a strong poloidal magnetic field~\citep[see e.g., Appendix A of][]{Etienne+15} such that $\max\bigl(P_{\rm mag}/P\bigr) = \max\bigl(b^{2}/2P\bigr) = 10^{-4}$, corresponding to $\max\sqrt{b^2}=5.05 \times 10^{15}~G$.

\begin{table}[!htb]
\begin{center}
\begin{tabular}{c|c} 
 \hline\hline
 \textbf{Parameter} & \textbf{Value} \\
 \hline
 \multicolumn{2}{c}{\textbf{Initial data}} \\
 \hline
 EoS & \sfho \\
 NS baryonic mass & 1.550\Msun \\
 NS gravitational mass & 1.348\Msun \\
 NS radius $R_*$ & 9.3\,km \\
 Initial separation & 45\,km \\
 $\max\sqrt{b^{\mu}b_{\mu}}$ & $5.05\times10^{15} G$ \\
 Number of tracer particles & $50000$ \\
 \hline
 \multicolumn{2}{c}{\textbf{Post-merger}} \\
 \hline
 BH dimensionless spin parameter, $\chi = a/M$ & 0.795 \\
 BH irreducible mass, $M_{\rm irr}$ & 2.444\Msun \\
 BH mass, $M_{\rm BH}$  & 2.726 \Msun \\
 Mass-weighted mean \ye $\left(r=60\,km, t\simeq 17 msec \right)$\  & $0.122$ \\
 [0.5ex] 
 \hline\hline
 \end{tabular}
 \caption{Simulation parameters for an equal-mass, magnetized BNS merger performed with \igm using a microphysical, finite-temperature EOS and a neutrino leakage scheme. }
  \label{table_mhd_torus}
  \end{center}
\end{table}

Where polar coordinates are more easily interpretable, we define such a system through the simple coordinate transformation
$r^2= (x^2 + y^2 + z^2)$, $\cos\theta = z/r$, $\tan\phi = y/x$, and the $xy$-plane is identical to the initial binary orbital plane. In these coordinates, the polar component of angular momentum is $u_\phi \equiv xu_y - yu_x$ becomes a conserved quantity when the spacetime is axisymmetric.

\section{Results} \label{sec:results}

\subsection{Original location of matter escaping the merger remnant}

\begin{figure}[t!]
\includegraphics[width=1.18\linewidth]{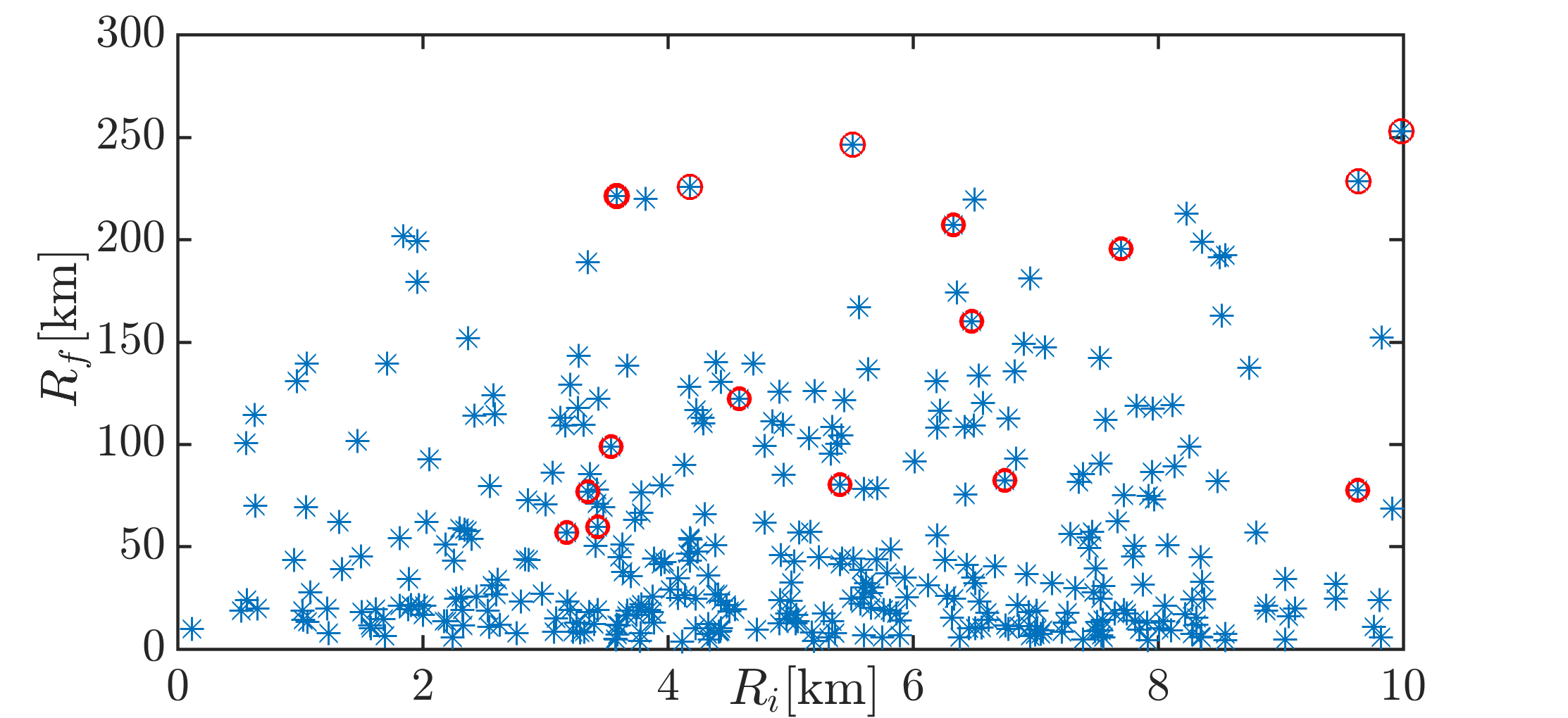}
\caption{The final distance $R_f$ (from the system center-of-mass) of the surviving tracer particles plotted against their initial distance $R_i$ from the center of their home neutron star. Red circles distinguish unbound particles.}
\label{fig:Rf_Ri}
\end{figure}

The most basic distinction between the tracer particles is their post-merger fate: unbound, bound, or confined within either a stable NS or a BH. We define unbound as having $hu_t < -1$, where $h$ is the specific enthalpy of the fluid surrounding the particle and $u_t$ is the particle's conserved energy. A bound particle is one with $hu_t > -1$, but located (in this case) outside the innermost stable circular orbit (ISCO) of the BH that forms (for this case, ${\simeq}3.7r_g = 14.8$~km from the center of the BH). We call the latter two categories ``survivors".

The simplest description of the particles' post-merger position is their radial distance from the system center-of-mass, $R_f$; here ``post-merger" means at the end of the simulation, ${\sim}18$~ms after its start. One might expect that particles initially placed near the surface of a NS would be more likely to ``survive" than particles whose starting position was deeper inside one of the stars. The relation between $R_f$ and the distance $R_i$ between each survivor particle and the center of its home NS is shown in Figure~\ref{fig:Rf_Ri}.
Only ${\simeq}950$ of the particles survive, 1.9\% of the original number. Roughly half are found outside $15r_g \simeq 60 \rm{km}$, and ${\sim}20\%$ outside $100 \rm{km}$. Almost {\it all} those outside $\rm 200\ km$ (${\simeq}8\%$ of the surviving particles) are unbound. Most importantly for our present purposes, {\it there is no correlation at all between $R_i$ and $R_f$} except for a slight (and statistically weak) tendency for the unbound particles to originate at larger $R_i$ than the bound particles.
Note that both $R_i$ and $R_f$ are coordinate distances defined by $\left[(x-x_c)^2 + (y-y_c)^2 + (z-z_c)^2\right]^{1/2}$, where $\left(x_c,y_c,z_c \right)$ are either the coordinates of the center-of-mass ($x_c=y_c=z_c=0$) or the coordinates of the nearest NS center.

\begin{figure}[t!]
\includegraphics[width=1.18\linewidth]{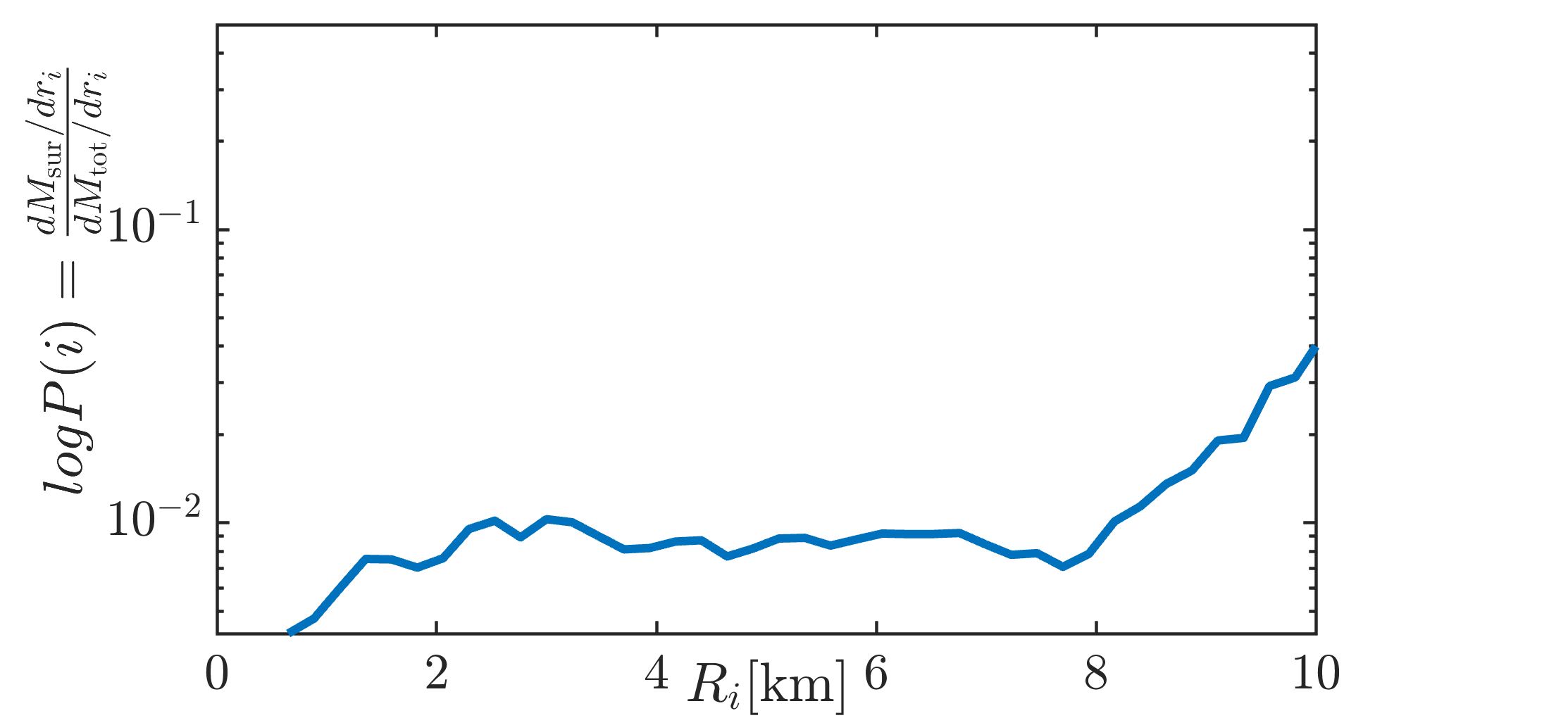}
\includegraphics[width=1.17\linewidth]{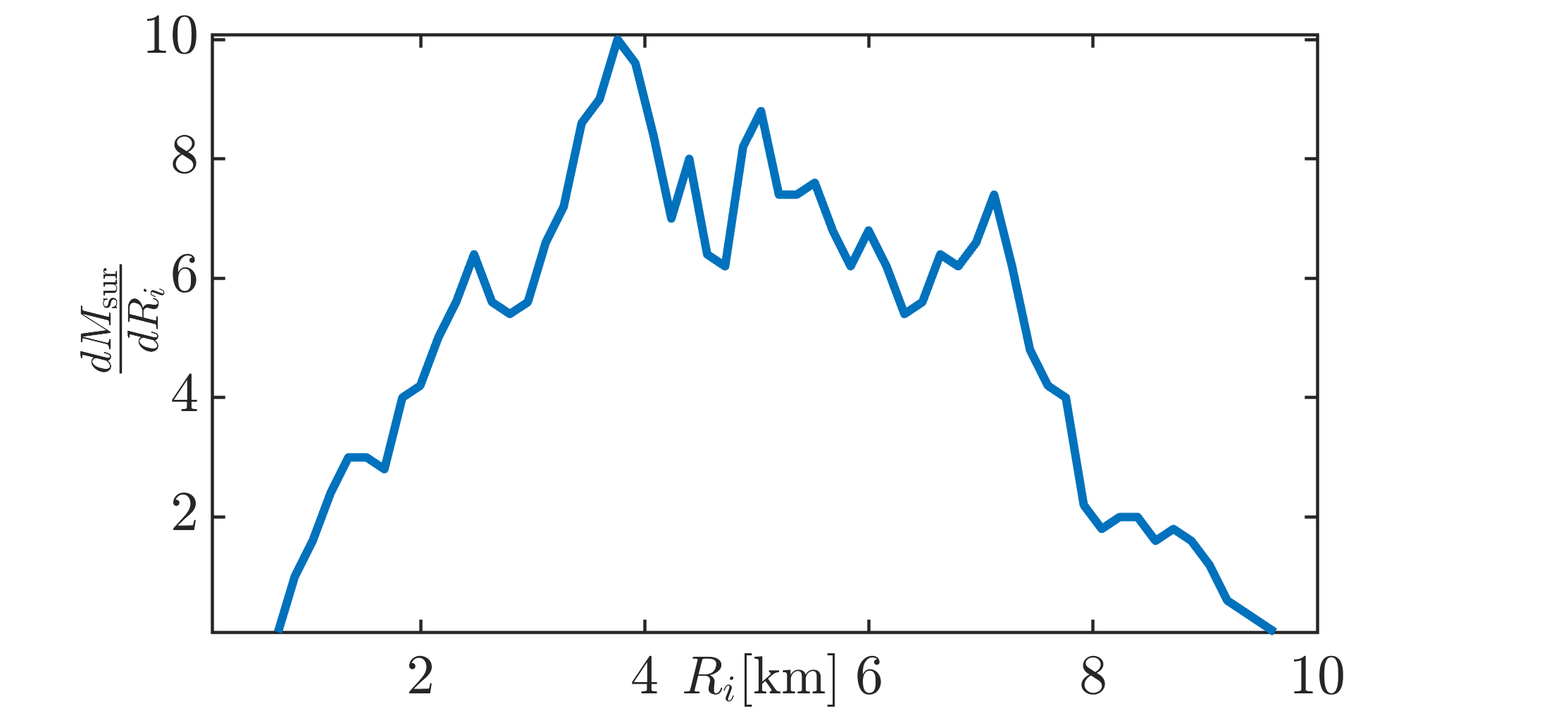}
\caption{Upper panel: The probability that a particle with initial radius $R_i$ survives black hole collapse. Lower panel: The probability density (in arbitrary units) that a surviving particle had initial radius $R_i$.}
\label{fig:Pdf_Ri_Rf}
\end{figure}

However, there is a sense in which particles whose initial home was at the surface of a NS {\it are} more likely to survive from the remnant. As shown by the upper panel of Figure~\ref{fig:Pdf_Ri_Rf}, the probability that a tracer particle escapes grow for $R_i > 8$~km, i.e., within ${\sim}1$~km of the surface. Compared to particles with smaller $R_i$ this probability is greater by ${\sim}3\text{--}4$. The reason why Figure~\ref{fig:Rf_Ri} shows no correlation is revealed by the bottom panel of Figure~\ref{fig:Pdf_Ri_Rf}: the probability density for the initial radius of a surviving particle is considerably larger over the range \mbox{$R_i \approx 2\text{--}8$~km} than for $R_i \gtrsim 8$~km. Even if a large fraction of {\it all} the surface particles escapes, while only a small fraction of interior particles do, there is so little mass near the surface compared to the mass in the interior that the great majority of surviving particles originate in the deep interior of a NS.

\subsection{Escape of debris from the NSs' interiors}

Although answering the first question we raised, this result prompts a second: how does the bulk of the debris find its way from the deep interior of a NS to an orbit well outside the merged NS or BH? Put another way, how is it that ${\sim}$2\% of the matter in the interior of a NS is somehow separated from its immediate neighbors and given enough energy and angular momentum to be released from the merged NS?

In order to approach this question, we begin by identifying the time from the beginning of the simulation at which this separation occurs.
We do so by choosing a small sample of surviving particles and then pairing each with the closest non-surviving particle in the initial state of the system. These were selected by finding survivors for which $R_* - 2$~km$\ < R_i < R_*$ in a single neutron star and were all located in different quadrants. We also constructed comparison samples varying these choices, but the results were quite similar.

\begin{figure}[t!]
\includegraphics[width=1.1\linewidth]{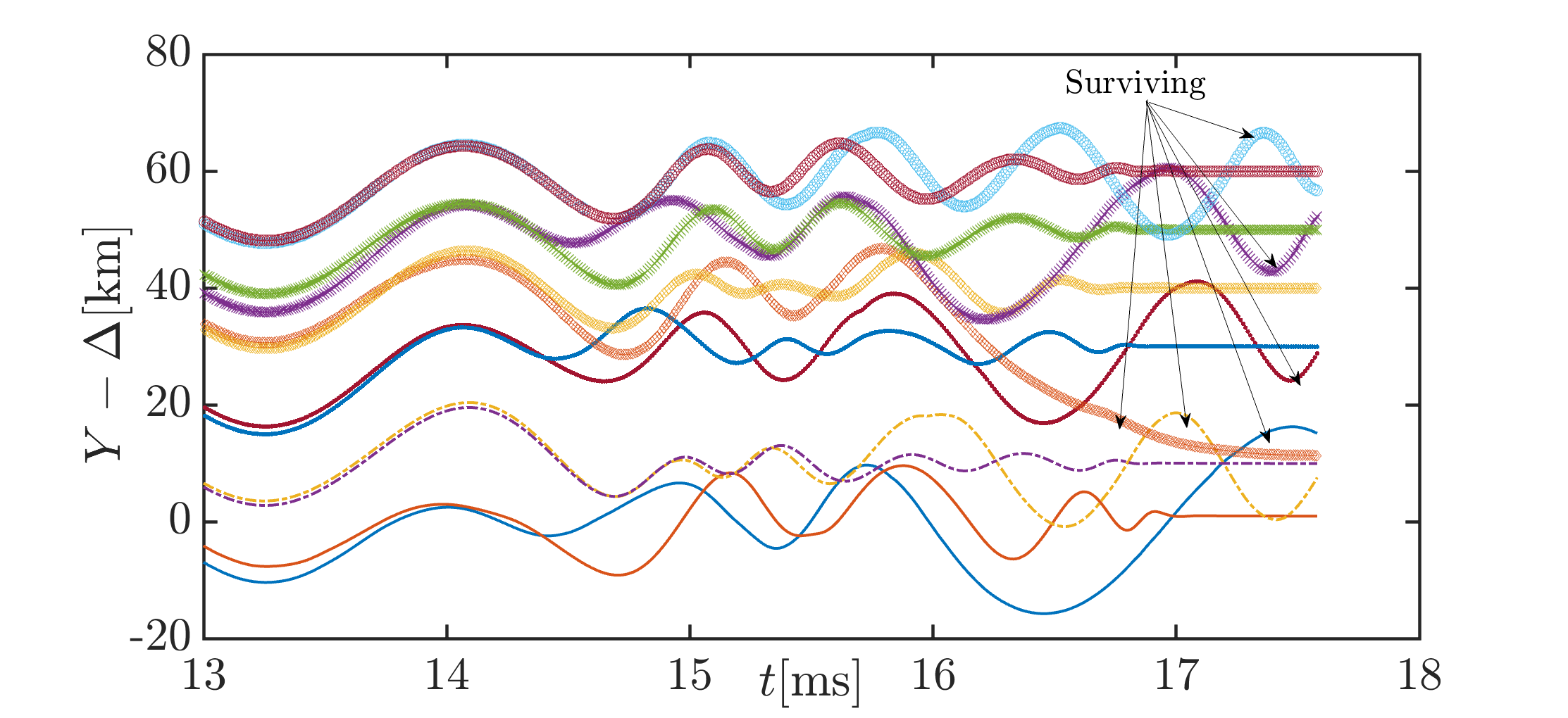}
\includegraphics[width=1.1\linewidth]{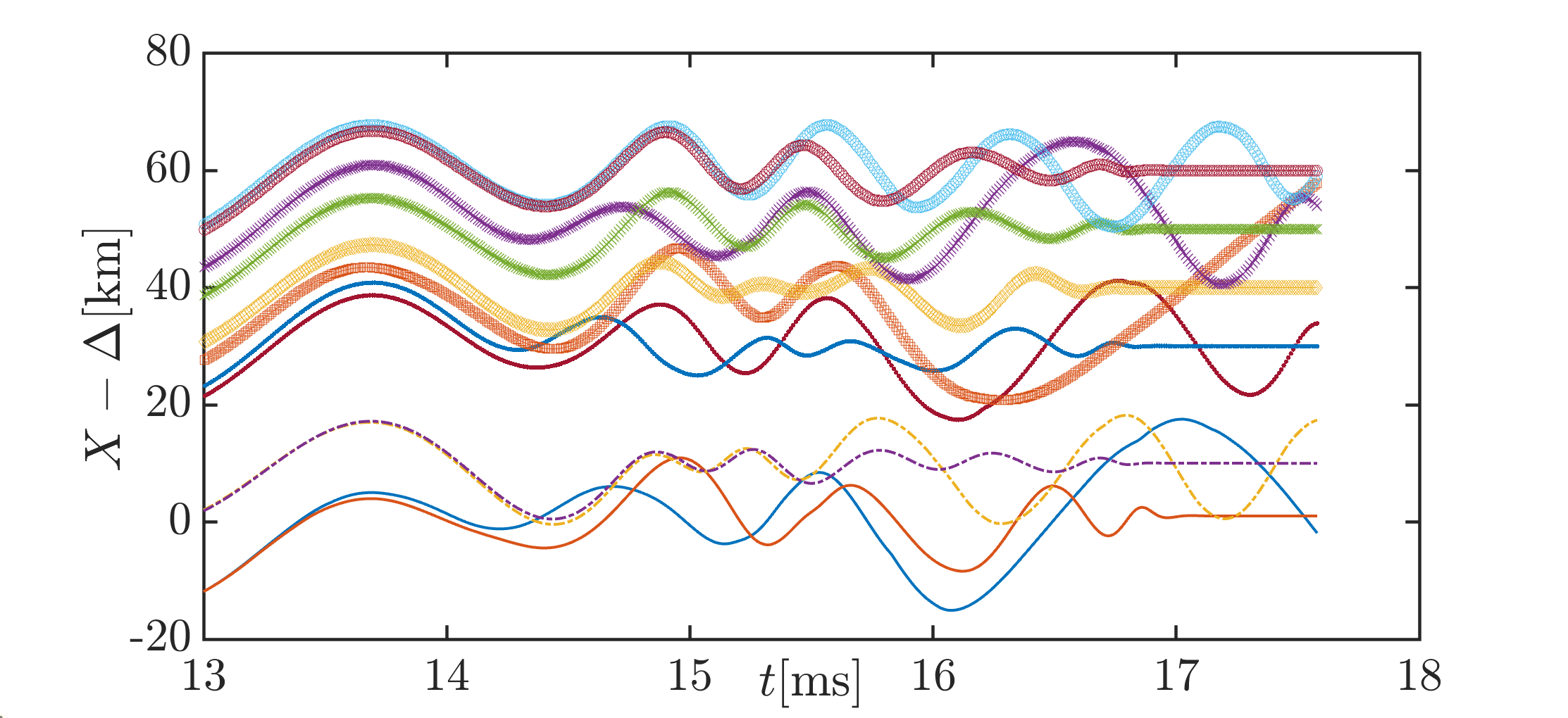}
\caption{The evolution of the separation between \emph{six} surviving particles and their initial close neighbors. The top panel shows how their \emph{y-}coordinates evolve, the bottom panel shows the evolution of their \emph{x-}coordinates. Each pair's starting location $(X, Y)$ is shifted by an arbitrary amount relative to the other pairs so that they can be visually distinguished. In some cases (e.g., the bottom two), the members of the pair start out so close together that their curves are superposed.}

\label{fig:Surv_nearst}
\end{figure}

In Figure~\ref{fig:Surv_nearst}, we display how the positions of these particles projected onto the $x$-$y$ (orbital) plane evolve from $t=13.0$~ms, 0.2~ms before the two stars first touch, to 17.62~ms, 1~ms after the black hole forms. The oscillations reflect the rotation of the NSs, first as they orbit one another, and later as they rotate as an asymmetric merged object. As this figure shows, four of the \emph{six} pairs begin to separate at ${\simeq}14.5\ \rm{ms}$, while the other two do not substantially part ways until ${\simeq}15.5$~ms. This small sample suggests the critical time for distinguishing surviving particles from captured particles is $\simeq$14.5--15.5~ms; as we will see shortly, this suggestion is confirmed when we examine the data more closely.

\begin{figure*}
  \begin{tabular}{ccc}
    \includegraphics[width=0.3\textwidth]{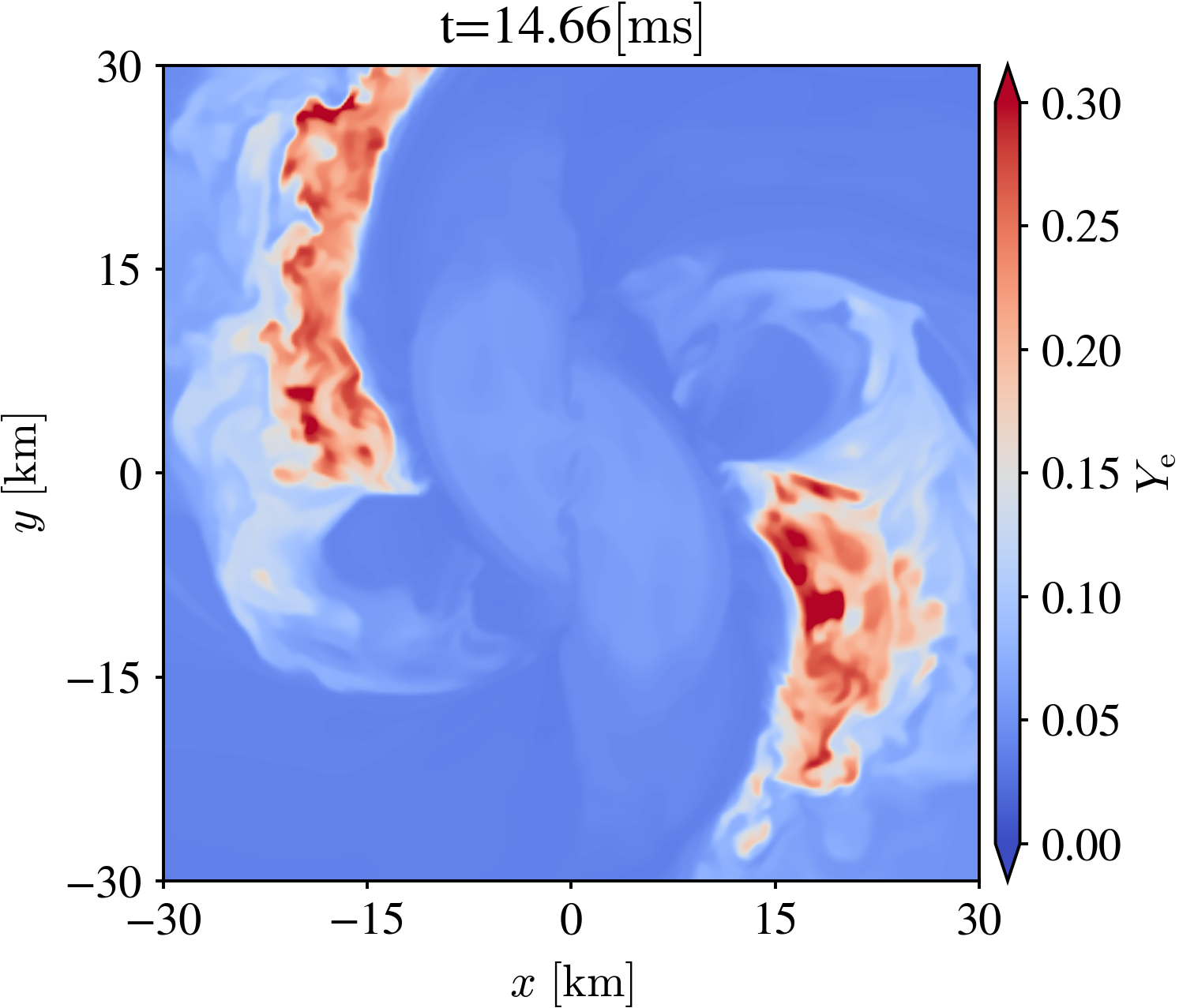} &%
    \includegraphics[width=0.3\textwidth]{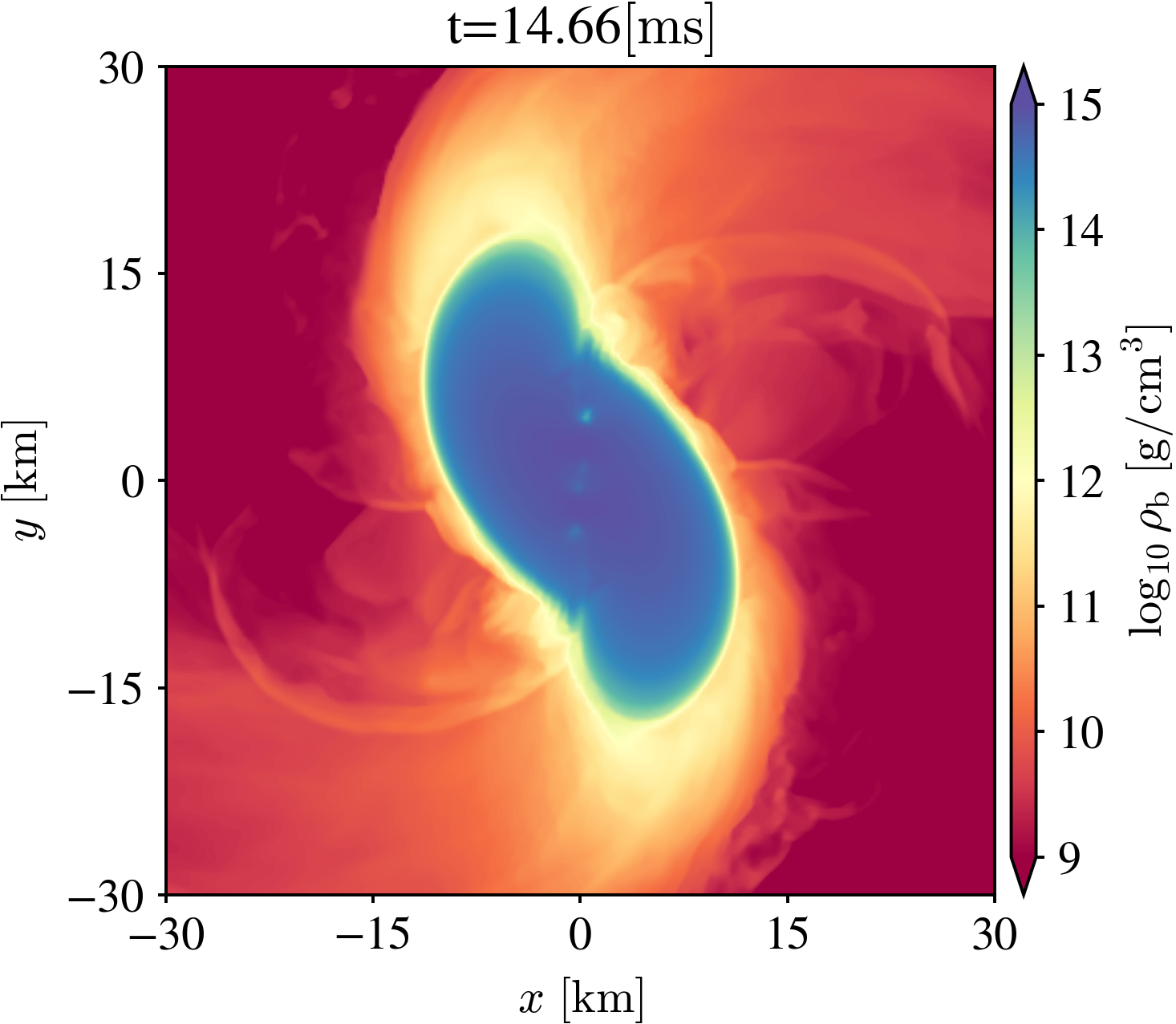} &%
    \includegraphics[width=0.3\textwidth]{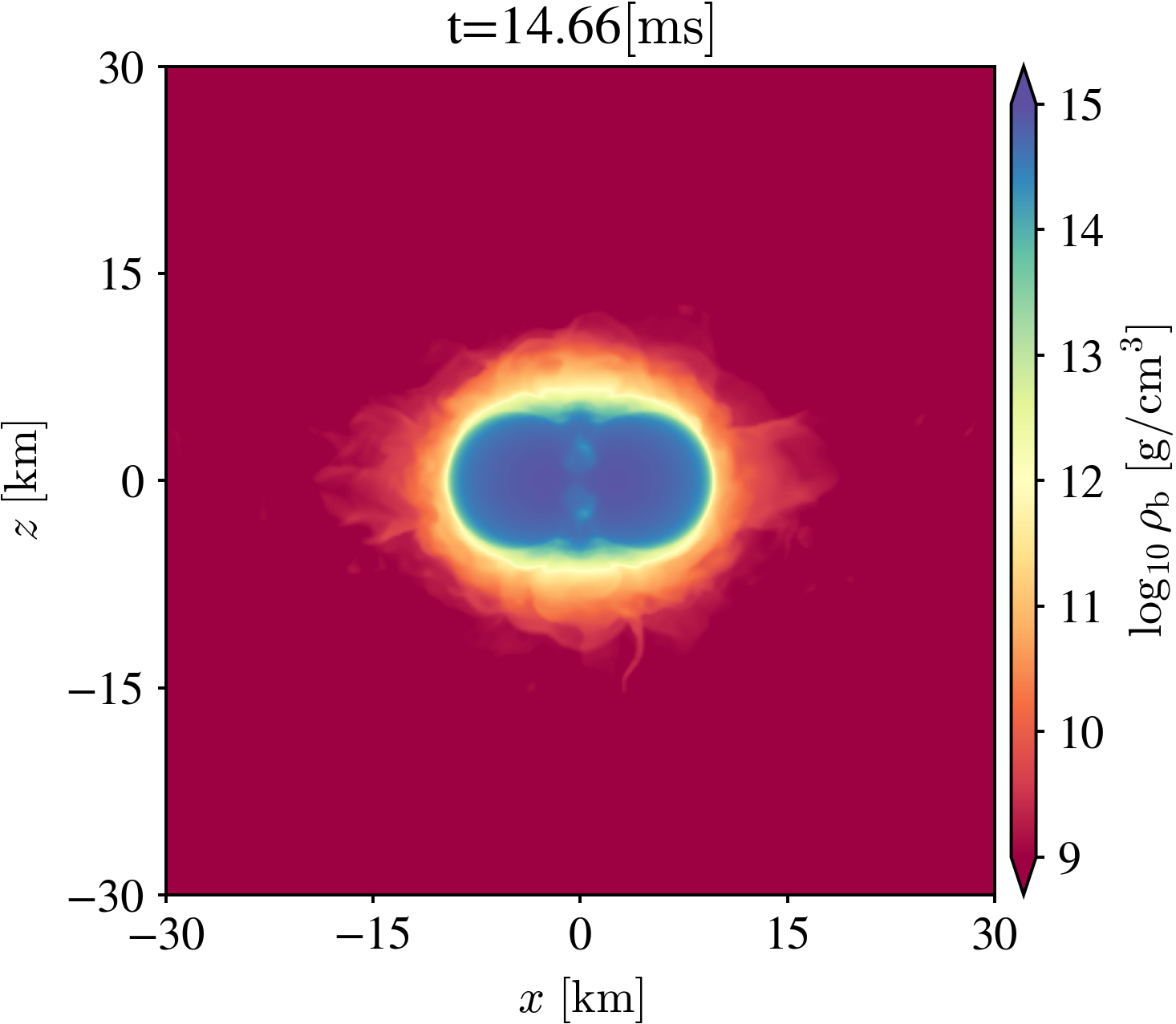} \\

    \includegraphics[width=0.3\textwidth]{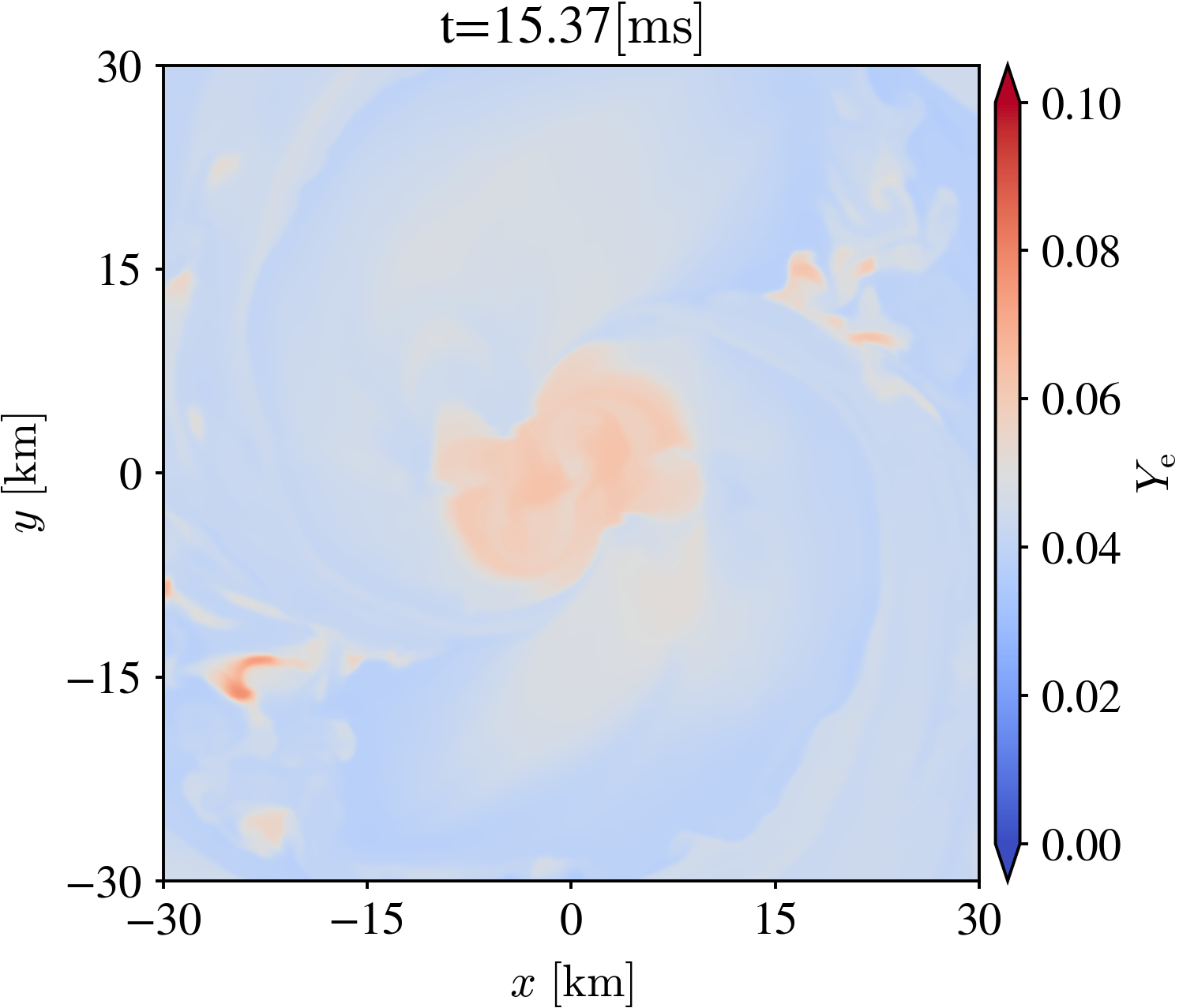} &%
    \includegraphics[width=0.3\textwidth]{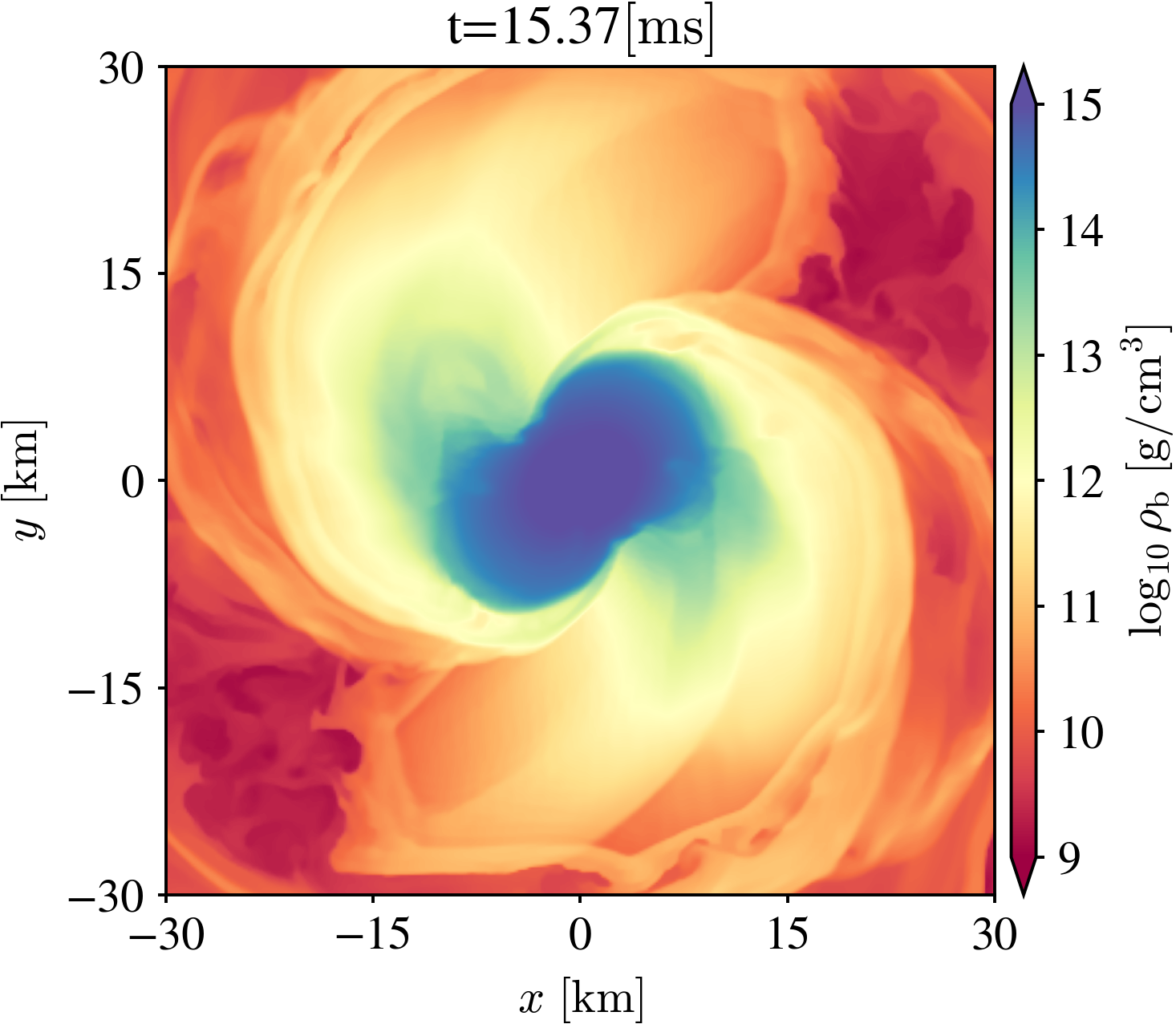} &%
    \includegraphics[width=0.3\textwidth]{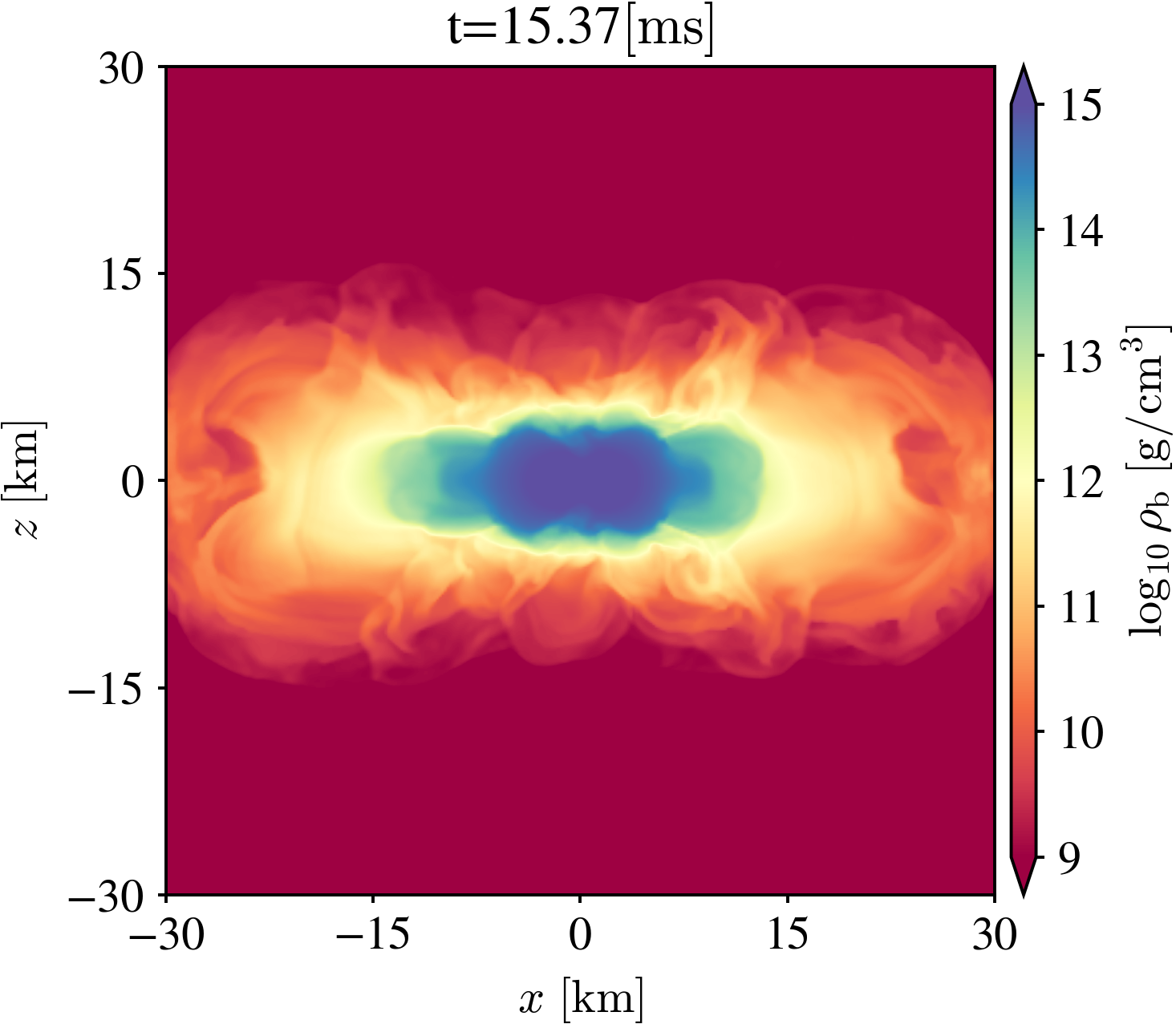} \\

    \includegraphics[width=0.3\textwidth]{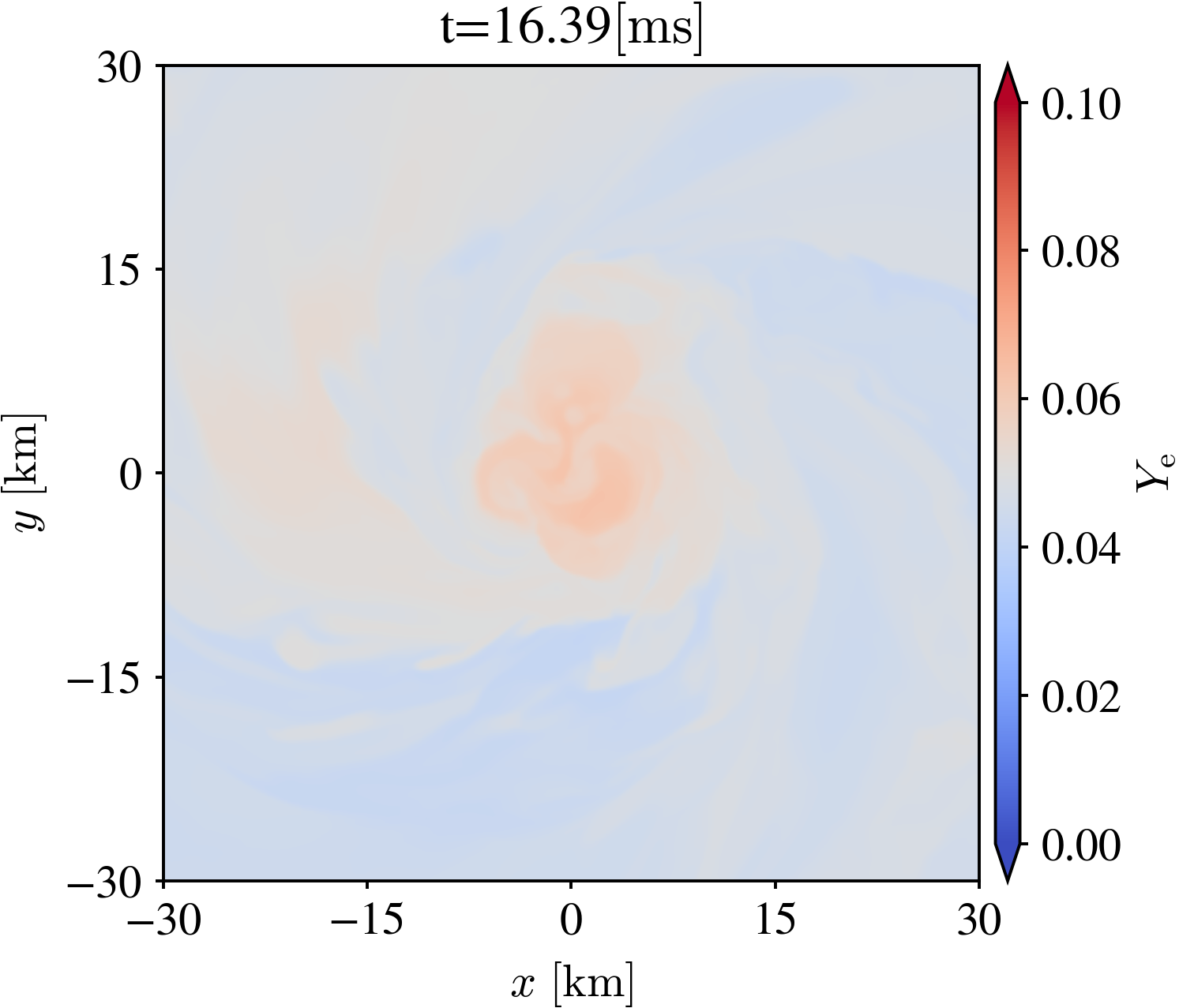} &%
    \includegraphics[width=0.3\textwidth]{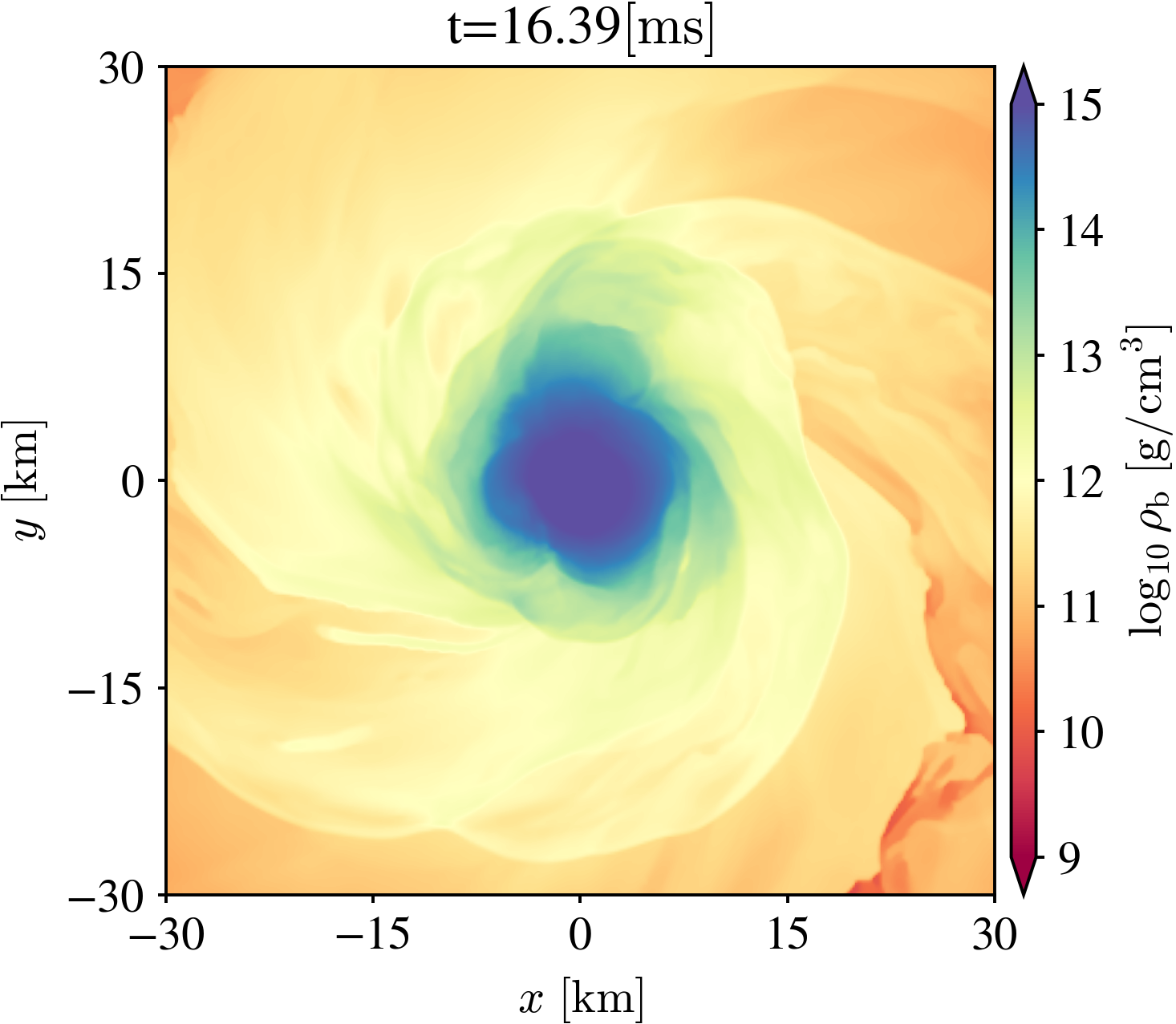} &%
    \includegraphics[width=0.3\textwidth]{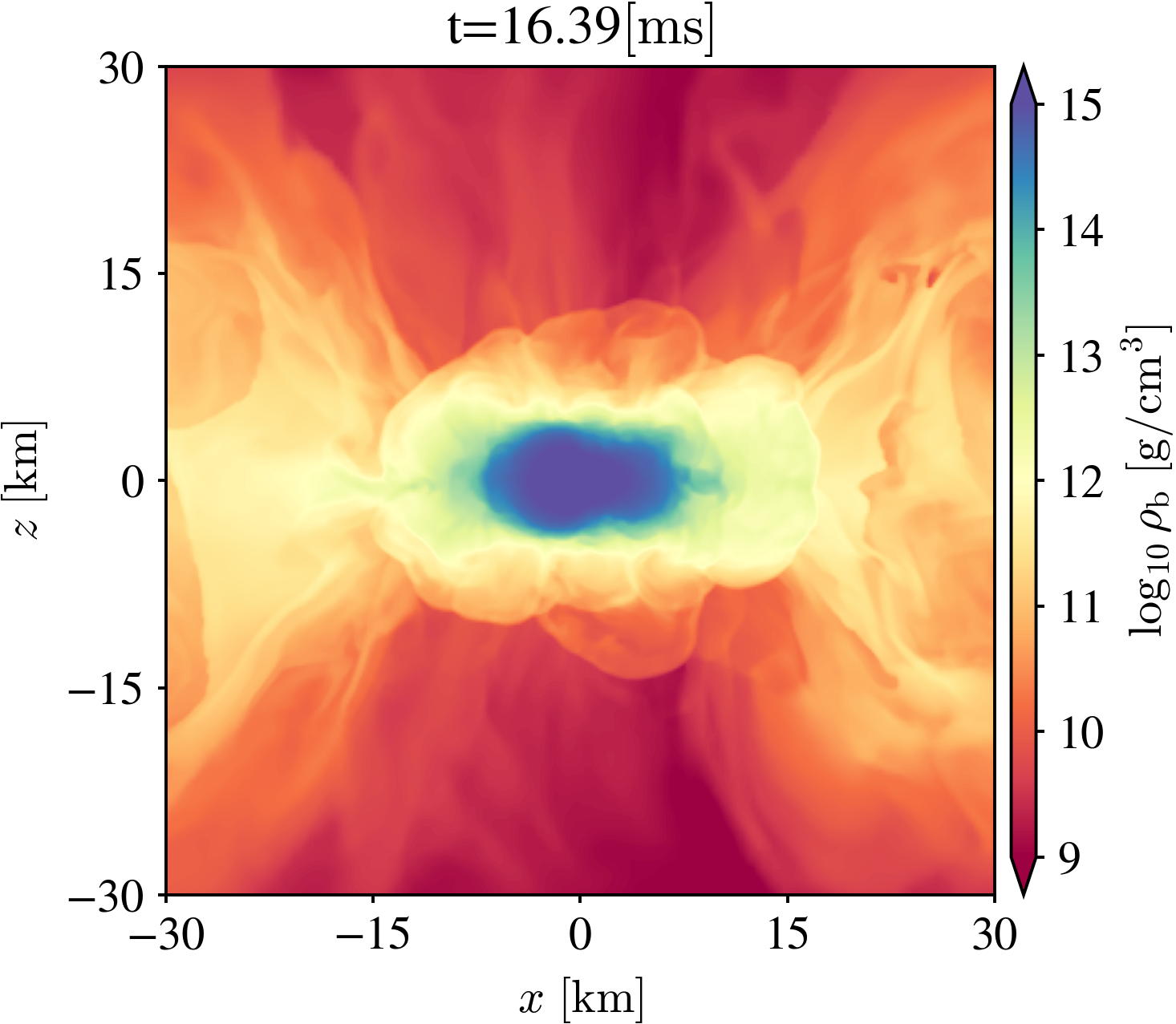} \\

    (a) \ye & (b) $\log\rho_{\rm xy}$ & (c) $\log\rho_{\rm xz}$
  \end{tabular}
  \caption{Evolution of the electron fraction, \emph{xy-}plane mass density, and \emph{xz-}plane mass density. $\ye$ is shown on a linear color scale (\emph{xy-}plane), the mass densities on a logarithmic scale. Each quantity is presented at times $\rm 14.66\ ms$, $\rm 15.29\ ms$, $\rm 16.31\ ms$, and $\rm 17.25\ ms$; the last is shortly after the collapse to a BH. Note that the colorscale for $\ye$ at $t=14.66$~ms is different from the colorscale at the other times.
  }
  \label{fig:Ye_Rho_B_xy}
\end{figure*}

To see what is happening during this crucial millisecond, we present images of the density and $\ye$ maps at that time (Fig.~\ref{fig:Ye_Rho_B_xy}). The first of the density images ($\rm t=14.66$~ms) shows the system ${\simeq}0.5$~ms after the star's first touch. Over the course of the following ${\sim}2$~ms, the configuration of the NSs changes from one in which the two stars remain partially distinguishable but are in contact over a sizable surface to one in which most of the stars' mass has taken on a much rounder shape, but a noticeable amount of gas has been expelled into spiral arms.

\begin{figure*}
  \begin{tabular}{ccc}
  \centering
   \includegraphics[width=0.37\textwidth]{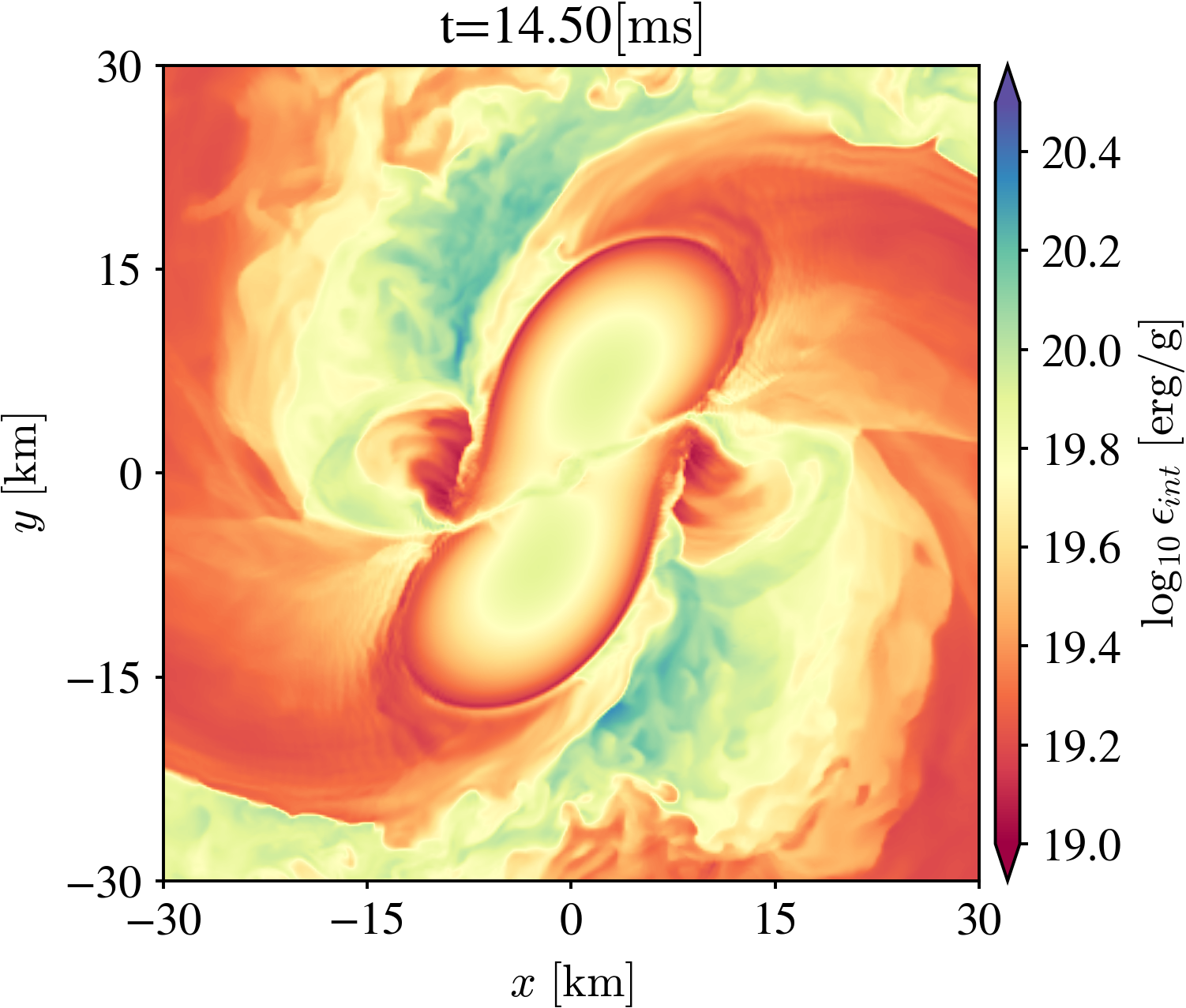} &
    \includegraphics[width=0.37\textwidth]{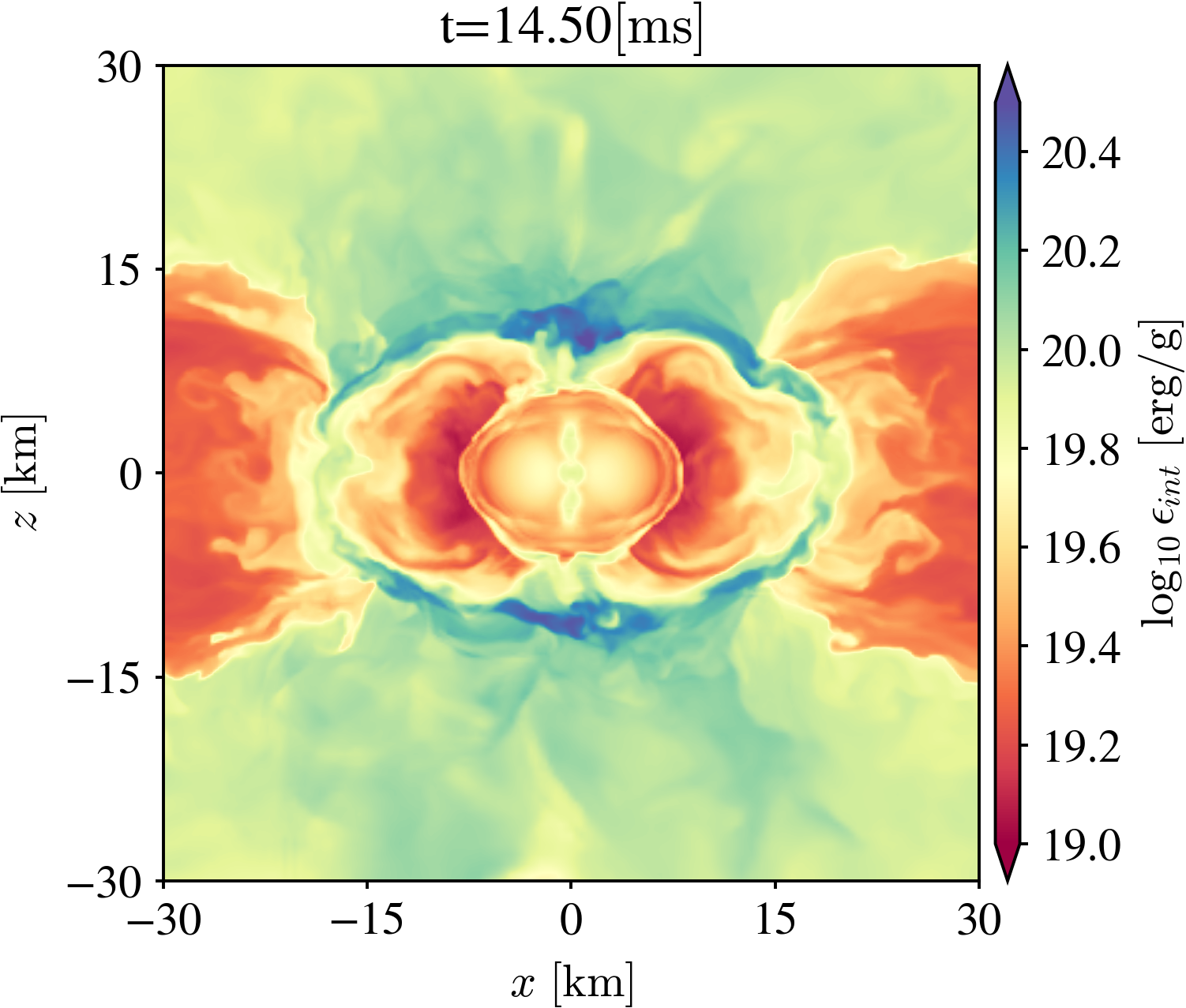} \\
    
    \includegraphics[width=0.37\textwidth]{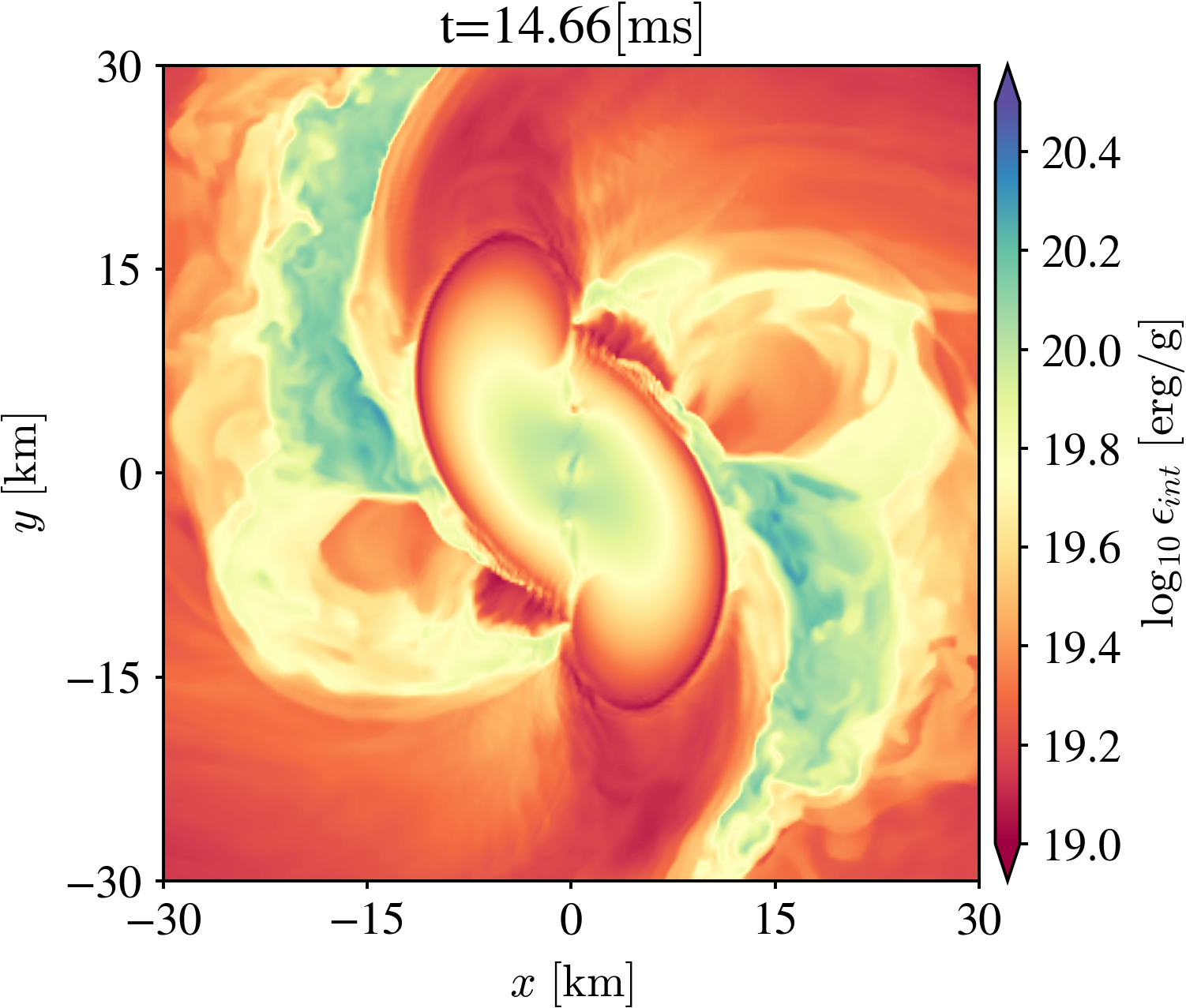} &
    \includegraphics[width=0.37\textwidth]{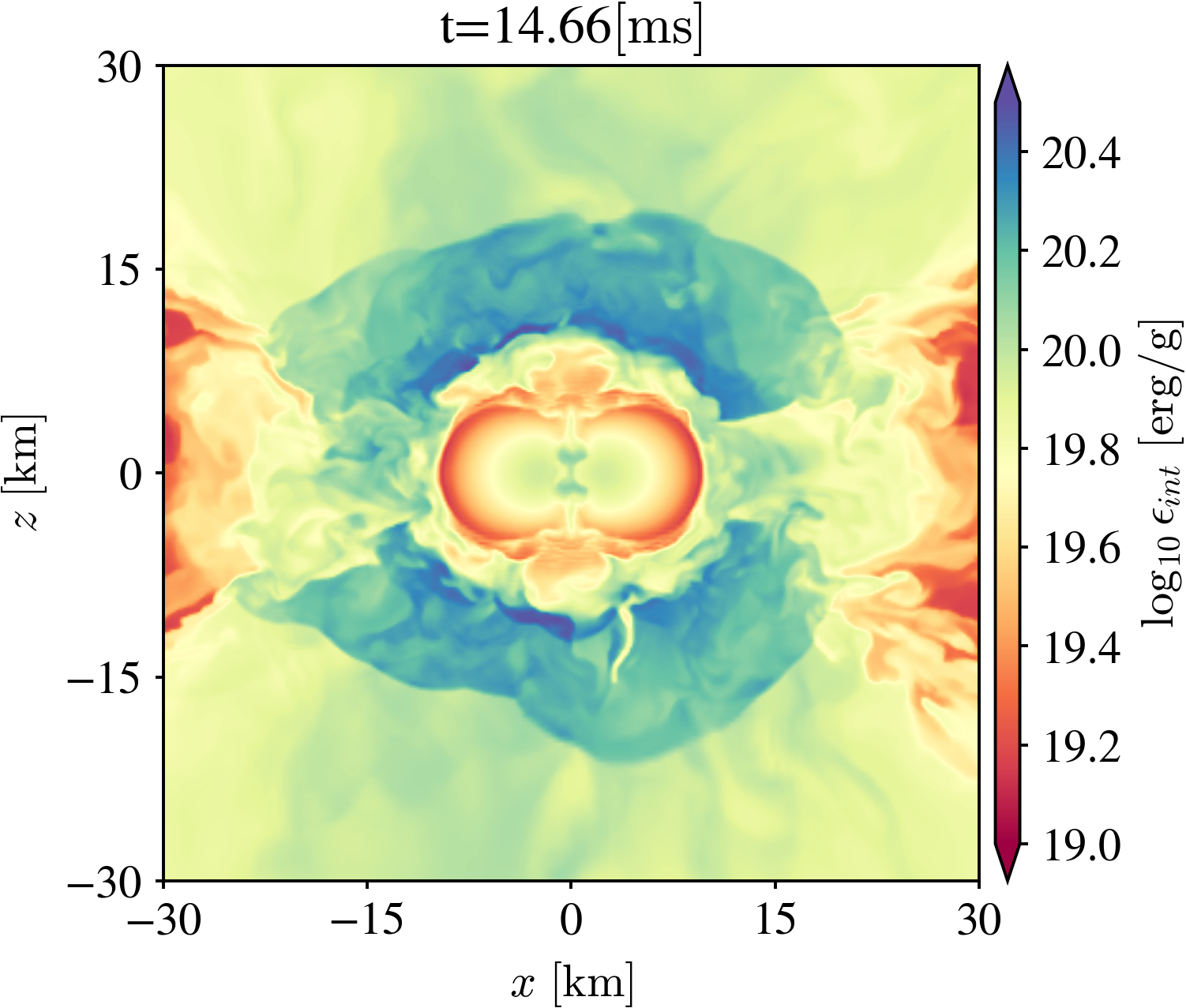} \\

    \includegraphics[width=0.37\textwidth]{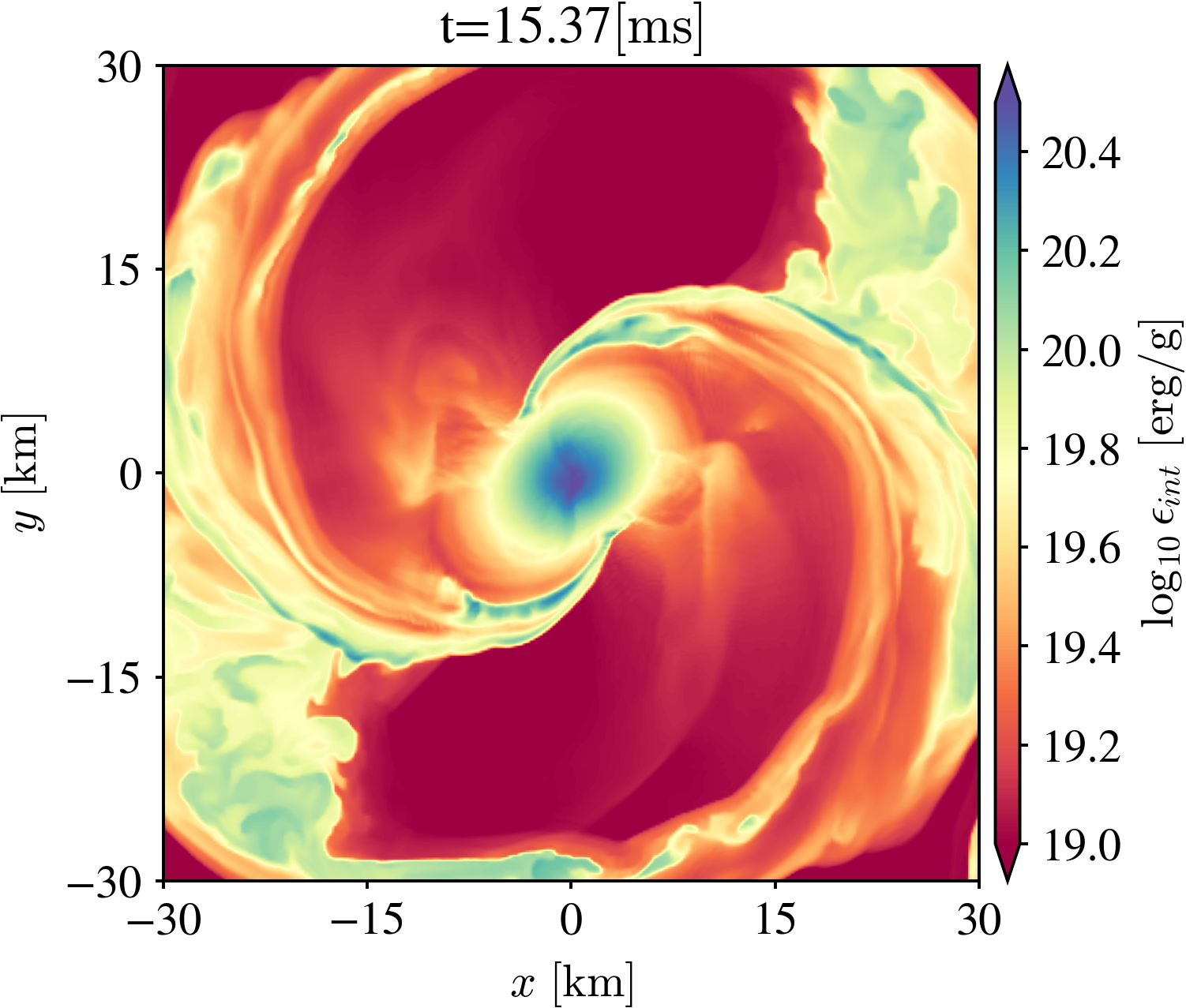} &
    \includegraphics[width=0.37\textwidth]{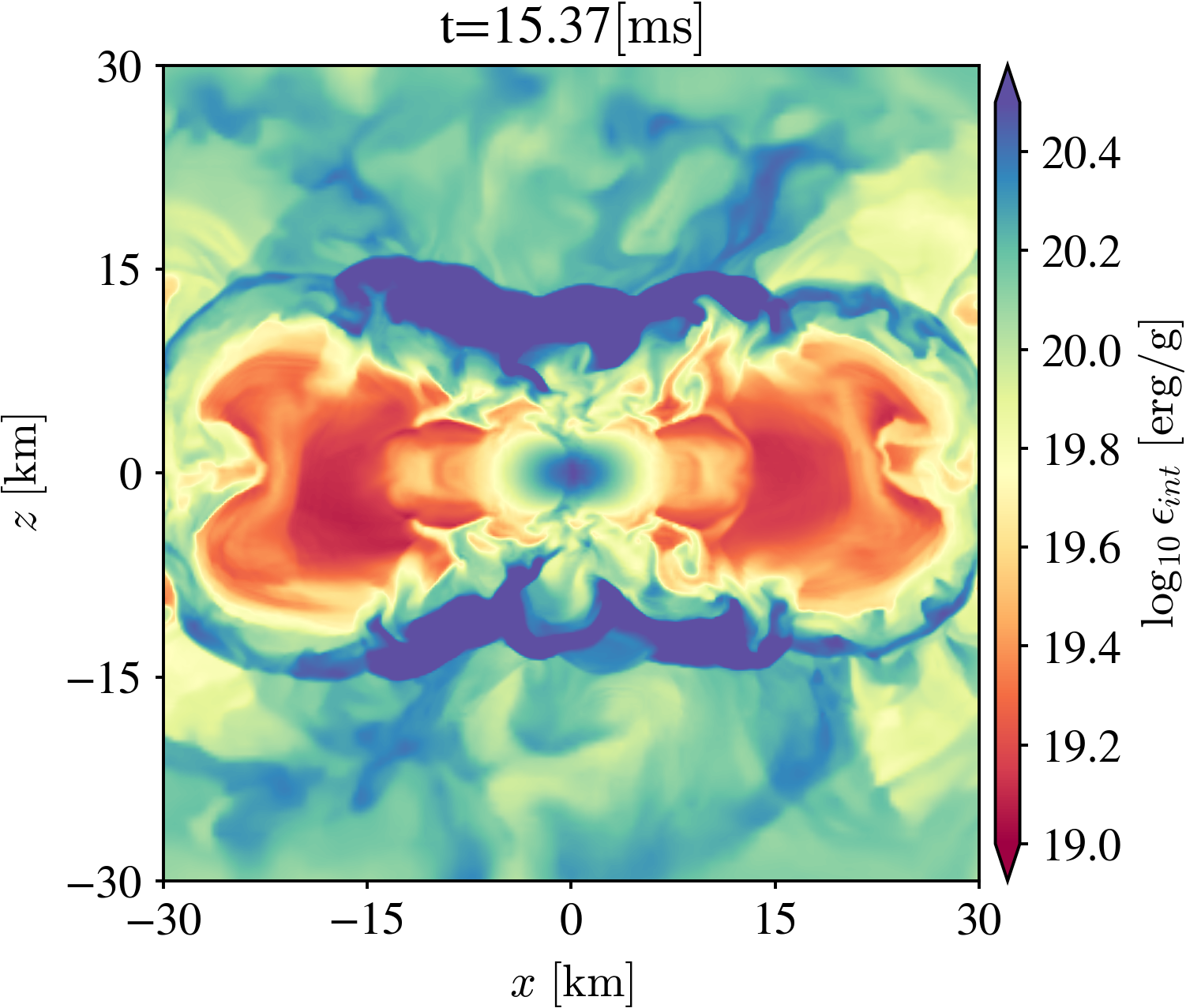} \\

    \includegraphics[width=0.37\textwidth]{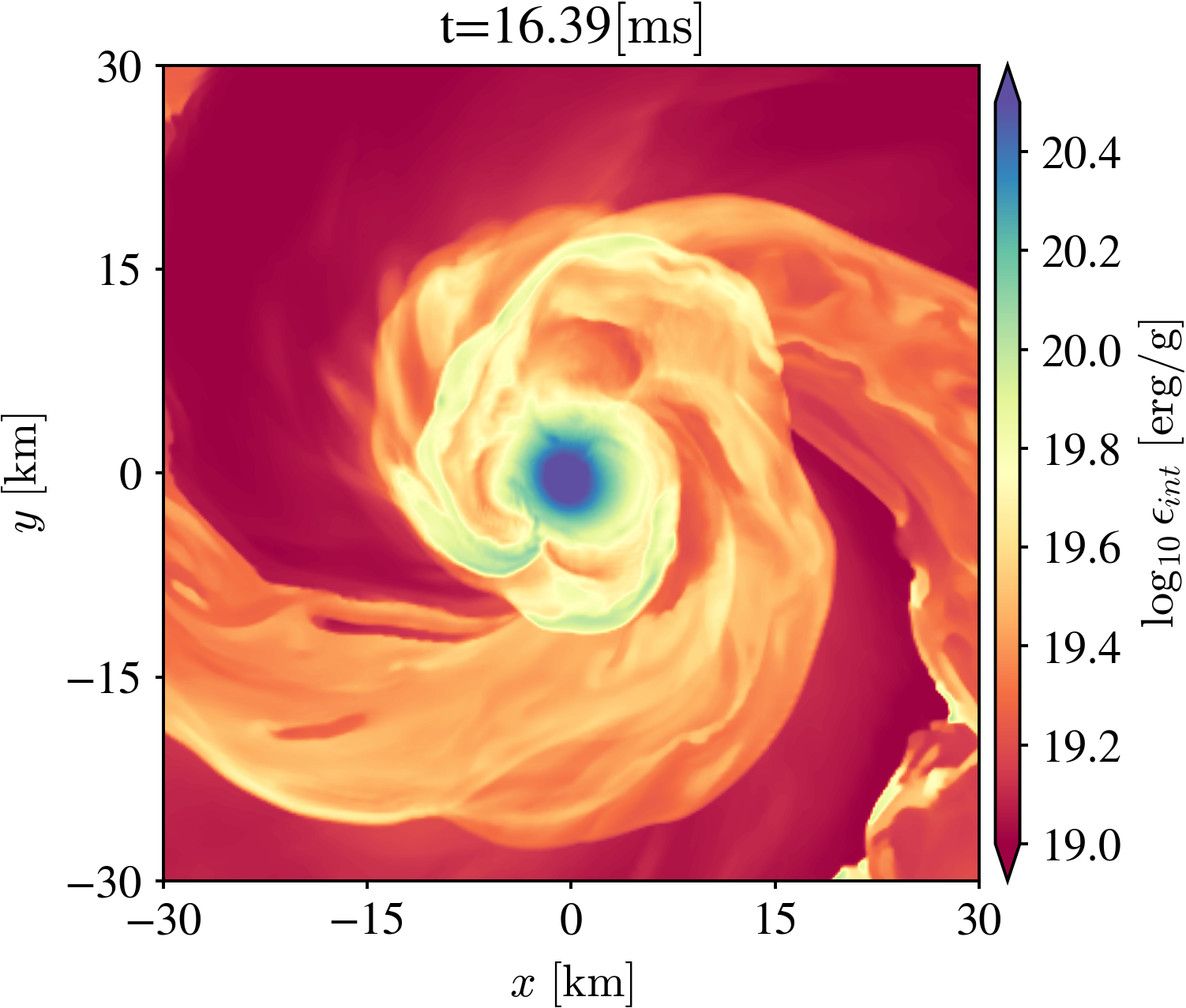} &
    \includegraphics[width=0.37\textwidth]{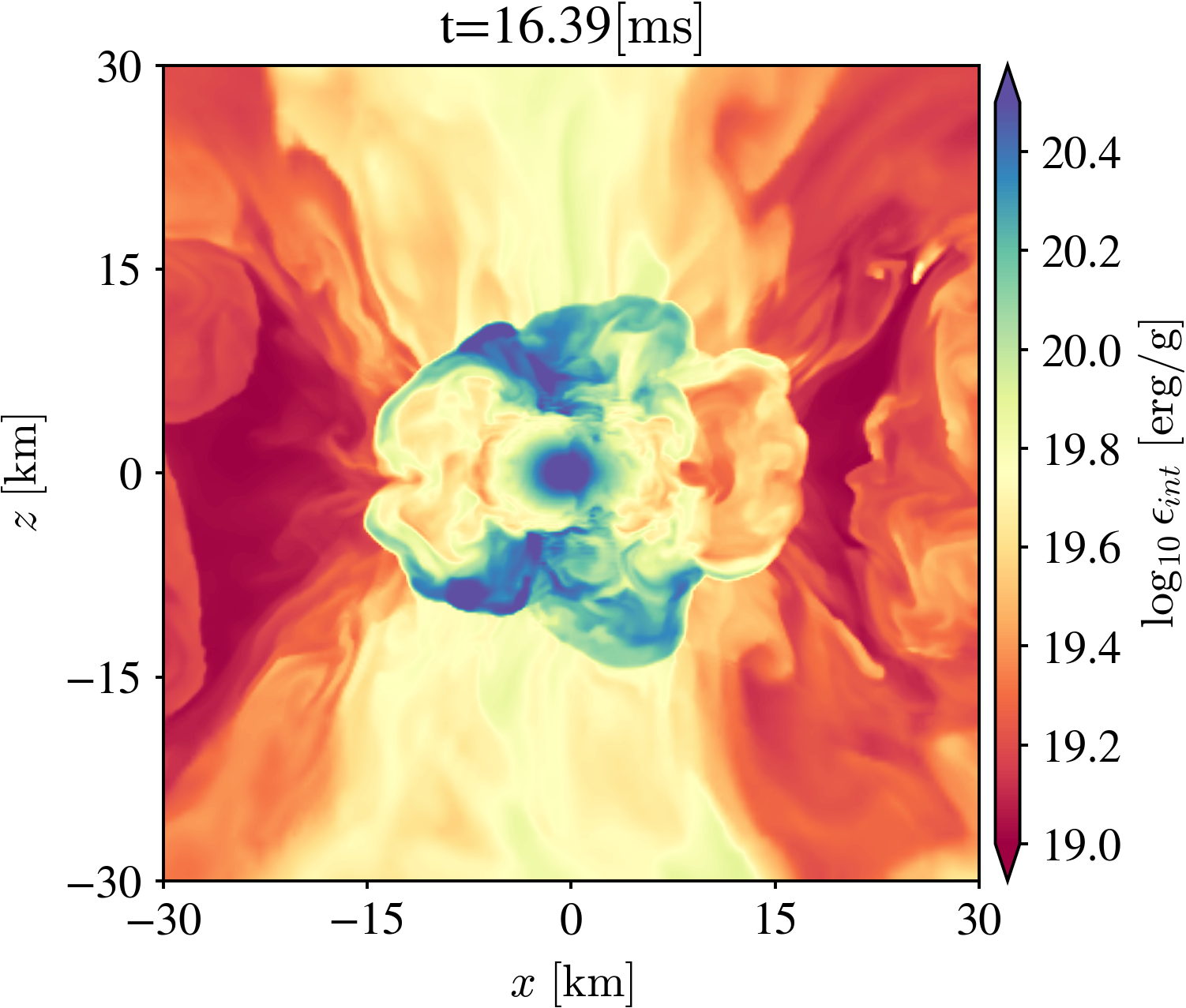} \\

    (a) $\log\epsilon_{\rm xy}$ & (b) $\log\epsilon_{\rm xz}$
  \end{tabular}
  \caption{Evolution of the specific internal energy $\epsilon_{\rm int}$ in the \emph{xy-} and \emph{xz-}planes, shown on a logarithmic scale at times $\rm 14.50\ ms$, $\rm 14.66\ ms$, $\rm 15.37\ ms$; $\rm 16.39\ ms$.}
  \label{fig:Eint_xy_xz}
\end{figure*}

Complementary images of the specific internal energy density are shown in Figure~\ref{fig:Eint_xy_xz}. For these, we begin at a slightly earlier time ($t=14.50$~ms) and follow it with the same three times shown in Fig.~\ref{fig:Ye_Rho_B_xy}. The internal energy images make shock locations easily visible. In particular, at the two earlier times, they show clearly the shock at the contact surface, while at the two later times they who how it weakens. They further point out shocks in surrounding gas driven by both polar outflow and the rotating asymmetric merged star \citep{Hotokezaka+13a}.

Thus, the moment at which the paths of escaping and captured particles diverge is the moment when the two stars are in direct contact, but have not yet combined their masses into a single (in this case, short-lived) NS. During the first part of this stage, the mass distribution is not only highly asymmetric; it also rotates around the center of mass, so spacetime is both asymmetric and time-dependent. However, during the latter part, the mass distribution becomes significantly more axisymmetric, weakening the azimuthal component of gravitational acceleration. Throughout this period, as shown by the numerous narrow regions of elevated internal energy per unit mass in Figure~\ref{fig:Eint_xy_xz}, there are many shocks, where kinetic energy is dissipated into heat, and deflection can exchange angular momentum between different fluid elements.

\begin{figure}[t!]
\includegraphics[width=1.05\linewidth]{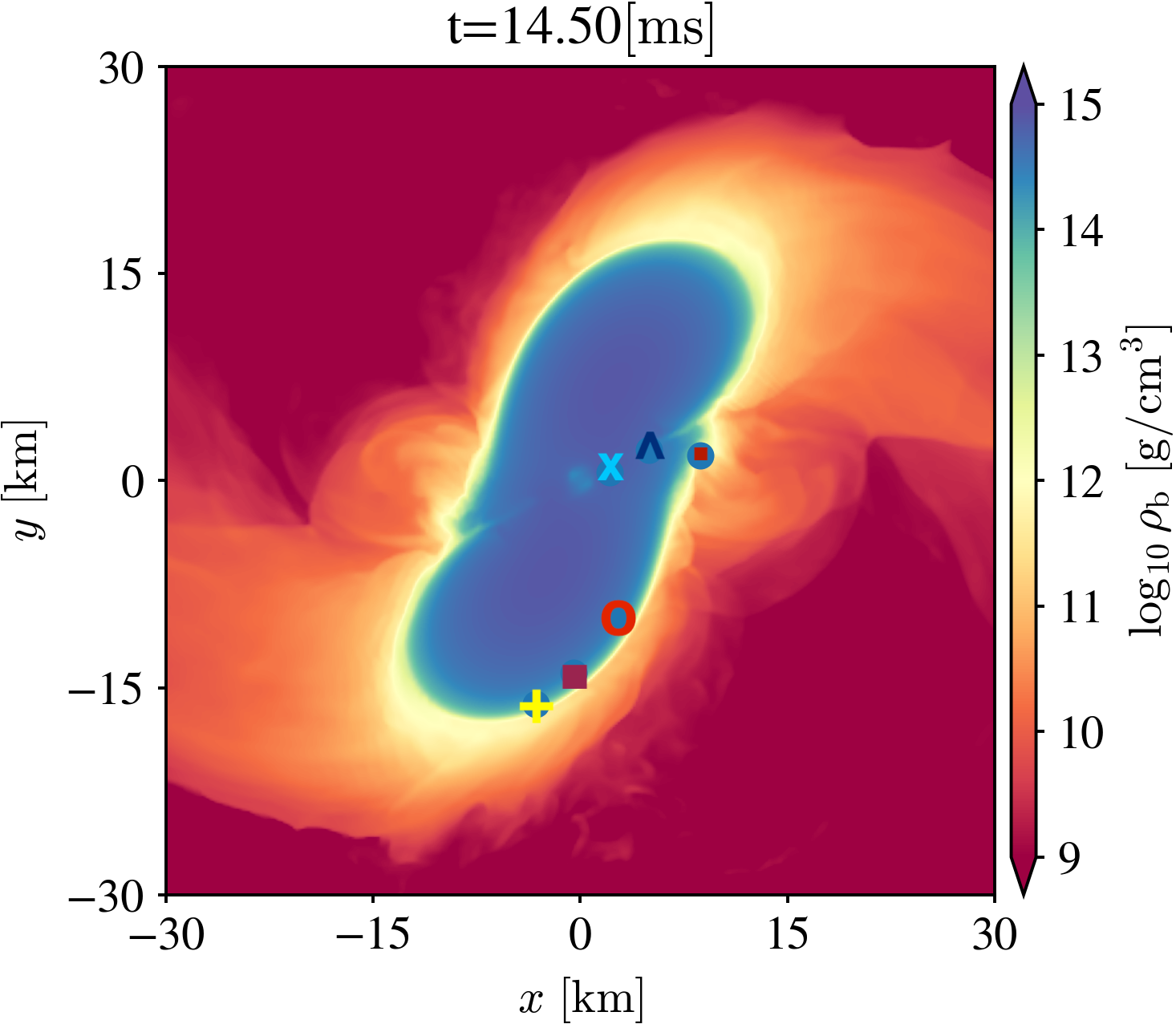}
\caption{The $x$-$y$ locations of the same \emph{six} surviving particles shown in Fig.~\ref{fig:Surv_nearst} superposed on the $t= 14.50 \rm{ms}$ density map. Their \emph{z-}coordinates are $z = 2.558, -2.225, -3.097, 2.182, 1.738, 3.504$~km for the particles shown as a plus-sign, a square, a circle, a dot, an 'X', and a caret, respectively.}

\label{fig:Surv_nearst_dens}
\end{figure}

To understand more deeply \emph{why} the surviving particles diverge from their neighboring captured particles, we now take a closer look at the \emph{six} surviving particles in our sample.
Figure~\ref{fig:Surv_nearst_dens} displays their $x$-$y$ positions at $t=14.50$~ms in relation to the fluid density in the orbital plane. Four of the \emph{six} particles' locations trace a curve in the \emph{x-y} plane running along the outer edge of the merged star; the other two are found along the ``seam-line" between the two merging stars. Their \emph{z-}coordinates inform us that these particles are typically $\sim (0.1 - 0.3)R_*$ out of the orbital plane, where $R_*$ is the radius of the original NSs.

\begin{figure}[t!]
\centering
\includegraphics[width=1.65\linewidth]{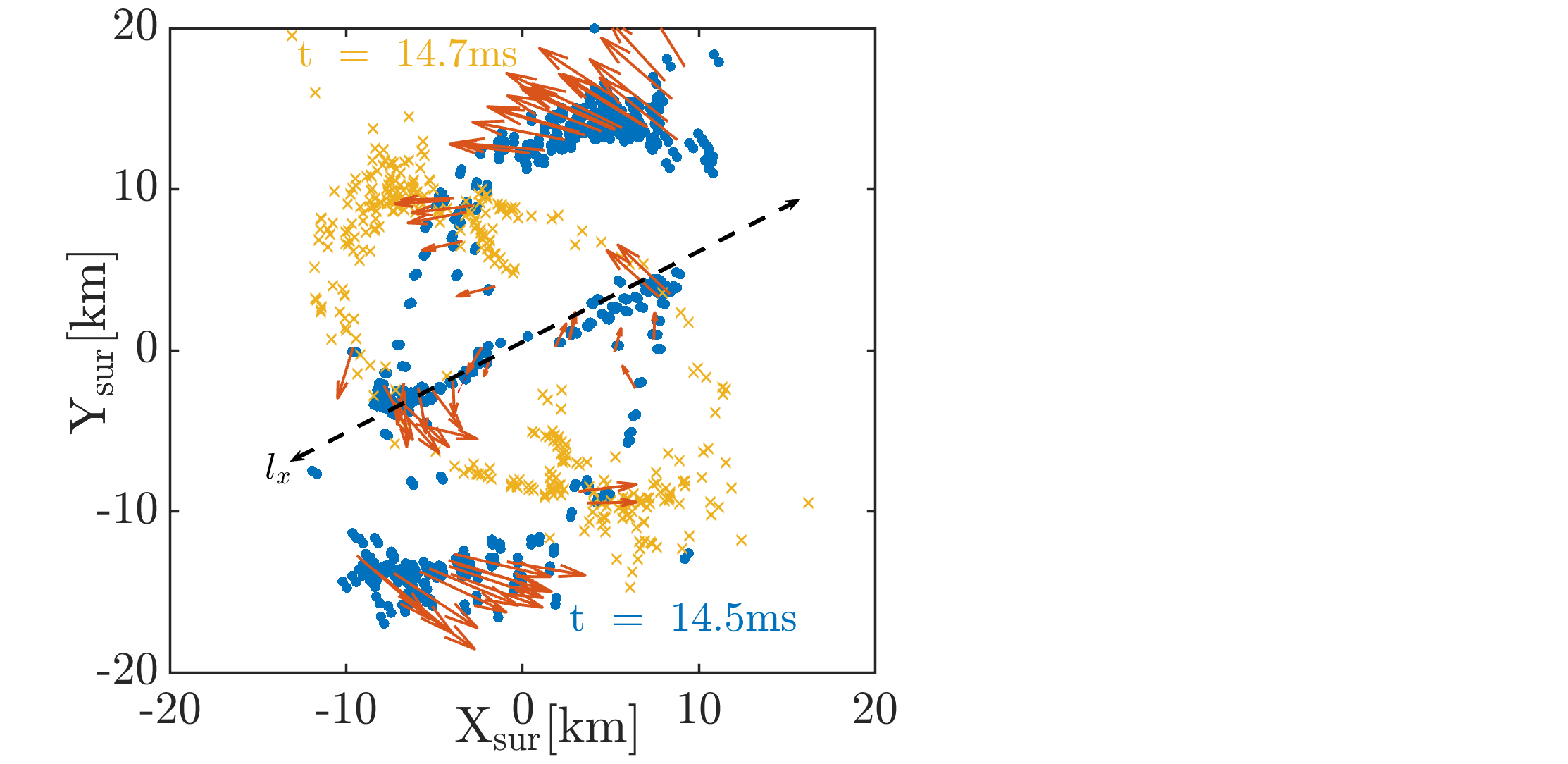}
\includegraphics[width=1.35\linewidth]{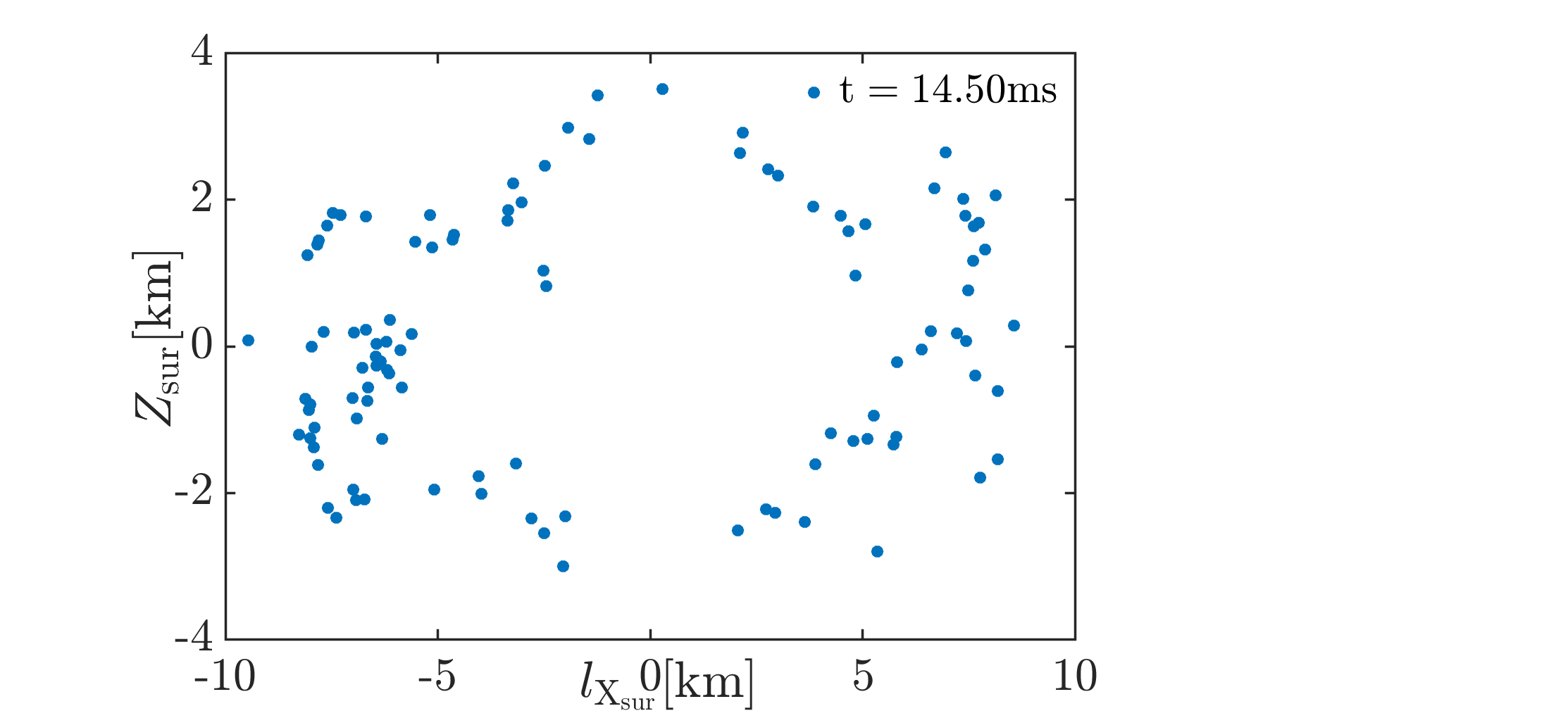}
\caption{(Top): Position in the \emph{xy-}plane of the surviving particles at $t = 14.506\ \rm{msec}$ (blue dots) and at $t = 14.705\ \rm{msec}$ (yellow crosses). The line labeled ``$l_x$" is fit to the line-like feature in the middle of the $\mathbf{``S"}$-shape at $t = 14.50\ \rm{msec}$. The red arrows illustrate the \emph{xy-}velocity of a random sample of surviving particles at $t=14.50$~ms; the arrows' lengths are proportional to the speeds of the particles' orbital plane motion. 
(Bottom): Position in the \emph{$l_x$z-}plane of the surviving particles in the line-like feature at $t=14.50$~ms. }
\label{fig:XY_XZ_t1466_lx_Z}
\end{figure}

To interpret this 3D geometry, we first plot the $x$-$y$ locations of \emph{all} the tracer particles at this same time (blue dots in the top panel of Fig.~\ref{fig:XY_XZ_t1466_lx_Z}). The four particles in our \emph{six}-member sample found along the edge of the merged star are all found in the lower hook of the $\mathbf{``S"}$-curve traced out by the entire ensemble of particles; the densest parts of both hooks coincide with the plumes of gas density visible in Figure~\ref{fig:Surv_nearst_dens}. The other two are in the central bar of the $\mathbf{``S"}$-curve, which coincides closely with the seam-line. It, therefore, appears that particles escape when they move outward through the seam region and then reach the surface. The curved portions of the $\mathbf{``S"}$-curve result from the time sequence at which the particles crossed through the surface of the merged NS. Note that the orbital plane velocities of the survivor particles are primarily azimuthal, and the radial gradient in their speeds indicates that they acquire greater angular momentum as they move outward.

Further information about vertical structure can be derived from examining the locations of the particles within the bar at $t=14.50$~ms. Their orbital-plane coordinates all lie along the linear feature in the $\mathbf{``S"}$, so that their 3D positions all lie in a vertical plane defined by the line feature and the $z-$axis (the simple geometry of this structure is likely due to our choice of an equal-mass merger). The bottom panel of Figure~\ref{fig:XY_XZ_t1466_lx_Z} shows their distribution in this plane. All of the particles lie a short distance inside the outline of the merged star, a figure that is flattened in the vertical direction (see the images in the third column of Fig.~\ref{fig:Ye_Rho_B_xy}). Moreover, they are weakly concentrated toward the orbital plane and tend to avoid the polar axis.

The evolution of this structure in time is illustrated by the difference between the figures outlined by the blue and gold dots in the upper panel of Figure~\ref{fig:XY_XZ_t1466_lx_Z}. At $t=14.50$~ms, numerous particles can be found in the plane of the inter-star seam, while the ``hooks", taken together, extend over $\sim 3\pi/2$ in azimuthal angle. Only 0.2~ms later, hardly any particles remain in the seam, while the hooks now cover all $2\pi$ in azimuth. Thus, the time period we have singled out as critical in terms of separating surviving particles from initially neighboring captured particles coincides with the time when the expulsion of matter from the seam region is completed.

\begin{figure}
\centering
\includegraphics[width=1.5\linewidth]{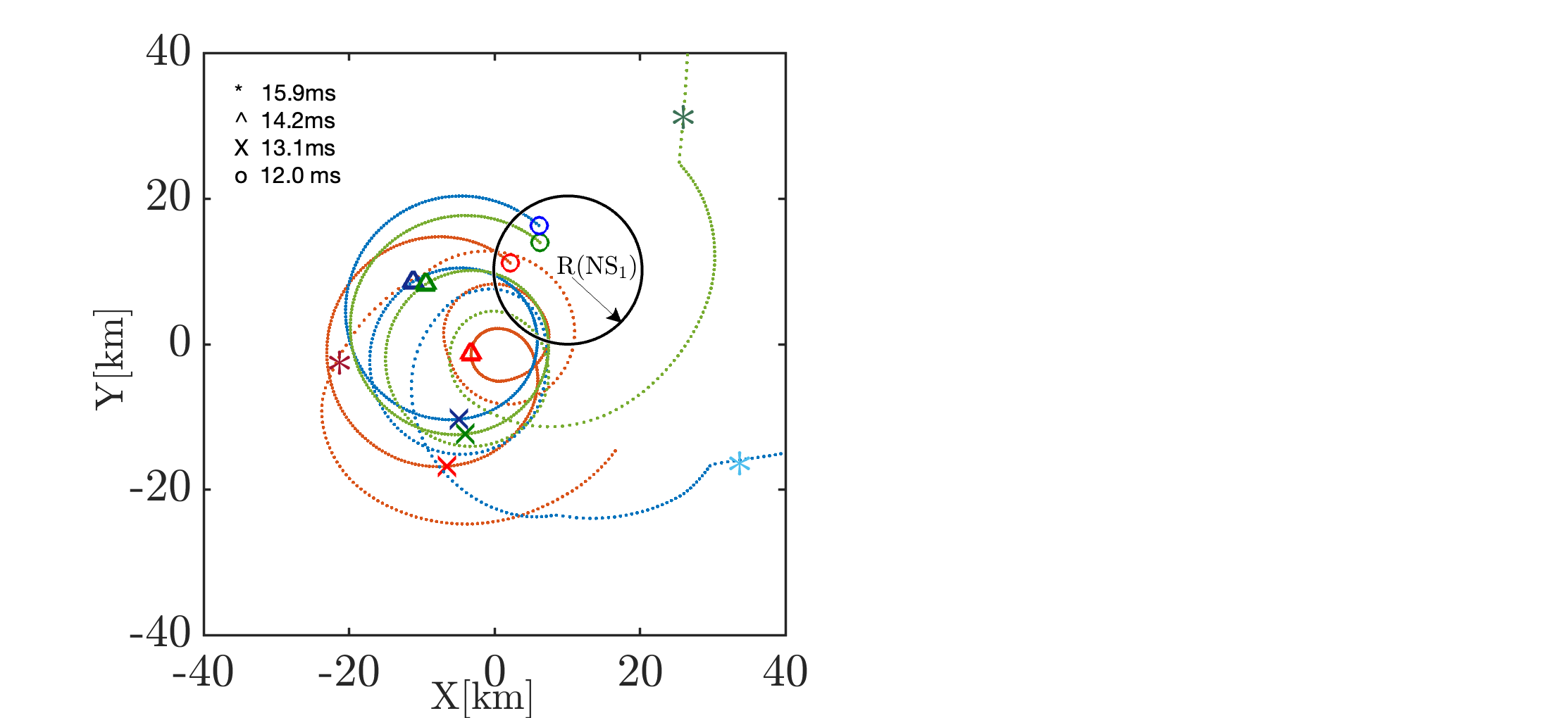}
\caption{Motion in $(x,y)$ of three surviving particles (blue, green, and red dotted paths) from $t=12$~ms to $t=16.6$~ms, when the star collapses to a BH. The black circle shows the outline of their parent star at $t=12$~ms. Cusps are the signature of shock deflection.}

\label{fig:XY_cm}
\end{figure}

The variety of paths these particles follow in order is illustrated in Figure~\ref{fig:XY_cm}, which shows three particle trajectories during the $\sim 2$~ms spanning the time of first contact between the stars. Just before contact, the tracer particles follow their star's orbit, but moving in and out relative to the system center-of-mass as the star rotates. Once the two stars come into contact (at $t = 13.2$~ms), the character of their motion changes: now they are part of the merged NS, partaking in its rotation around the system center of mass. However, the axial fluid motions within the contact surface permit particles to travel both outward and inward; the red path in Figure~\ref{fig:XY_cm} is an example of the latter option. Large amplitude motions within this zone are expected because the strong shear at the boundary between the two stars excites turbulence through the Kelvin-Helmholtz instability \citep{Rasio_Shapiro_99, Kiuchi+14, AguileraMiret+22}. By $t=14.7$~ms, only a few particles remain in the seam-plane. Instead, the particles destined to survive that had been inside the merger remnant at 14.5~ms are almost all found in a shell surrounding the remnant. The survivor particles that at 14.5~ms had been in the hooks of the ``S" are now in either of two places: roughly half are in the shell surrounding the remnant, many of them having moved {\it inward} to do so; the other half are outside the boundary of the figure.

\begin{figure*}
\includegraphics[width=0.98\linewidth]{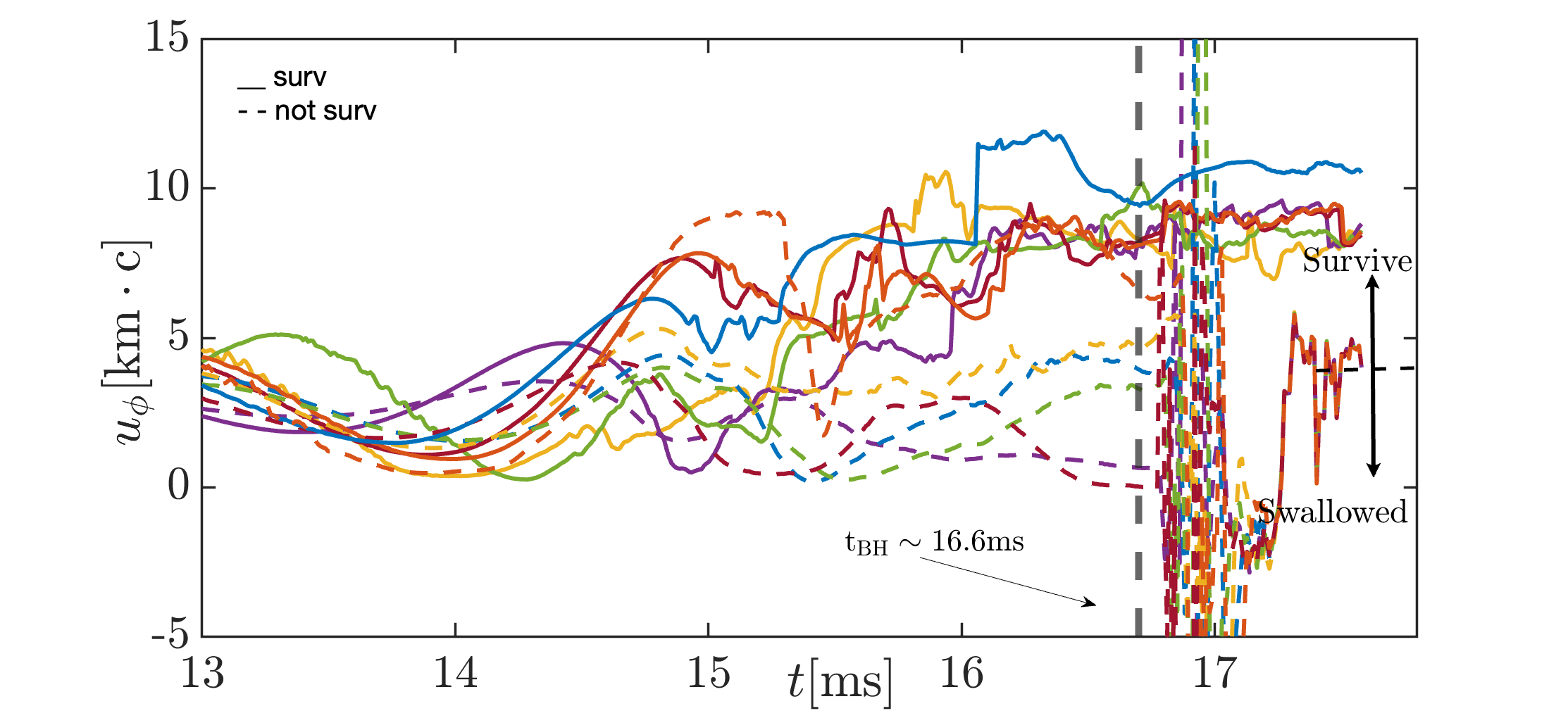}
\caption{Time-dependence of the angular momentum $u_\phi$ for our sample of \emph{six} pairs of particles. Survivors are shown with solid lines, and captured particles with dashed lines; matching colors identify the pairs. The time of collapse to a BH, $16.6$~ms is marked with a gray dashed vertical line; for captured particles after that time, the physical significance of $u_\phi$ is unclear.}

\label{fig:uphi_compare}
\end{figure*}
\subsection{Separation of surviving particles from neighboring particles}

The surviving particles are $\simeq 2\%$ of the NSs' mass; as can be seen in Figure~\ref{fig:XY_XZ_t1466_lx_Z}, $\simeq 90\%$ have reached positions just outside the star by $\sim 1$~ms after contact. From the data illustrated in Figure~\ref{fig:Surv_nearst}, it is clear that the captured particles that began close to the survivors are still close to them at this time. It immediately follows that the distinction between survivors and the captured matter is accomplished in two stages. First, a few percent of the total mass of NSs is conveyed to locations outside the stars by motion through the contact surface. Second, when this matter reaches locations immediately outside the merged star, local effects divide this mass into two subsets, one that escapes, the other remaining with the merged star.

How this distinction is made is portrayed in Figure~\ref{fig:uphi_compare}. For the first $\sim 2$~ms shown, the angular momenta of the surviving particles are very similar to those of their matching captured particles; both oscillate as a result of the NSs' rotation. At around $\rm 14-15\ ms$, the angular momentum of surviving particles begins to rise (as shown by the arrows in Fig.~\ref{fig:XY_XZ_t1466_lx_Z}), while the angular momentum of the neighboring particles that remain with the star for the long run do not change appreciably. This growth in survivor particle angular momentum finishes shortly before collapse to a BH ($t=16.6$~ms). By then, the surviving particles' angular momentum is $\sim 2 - 10\times$ the angular momentum of the captured particles that started out from locations very near them.

\begin{figure}
\centering
\includegraphics[width=0.99\linewidth]{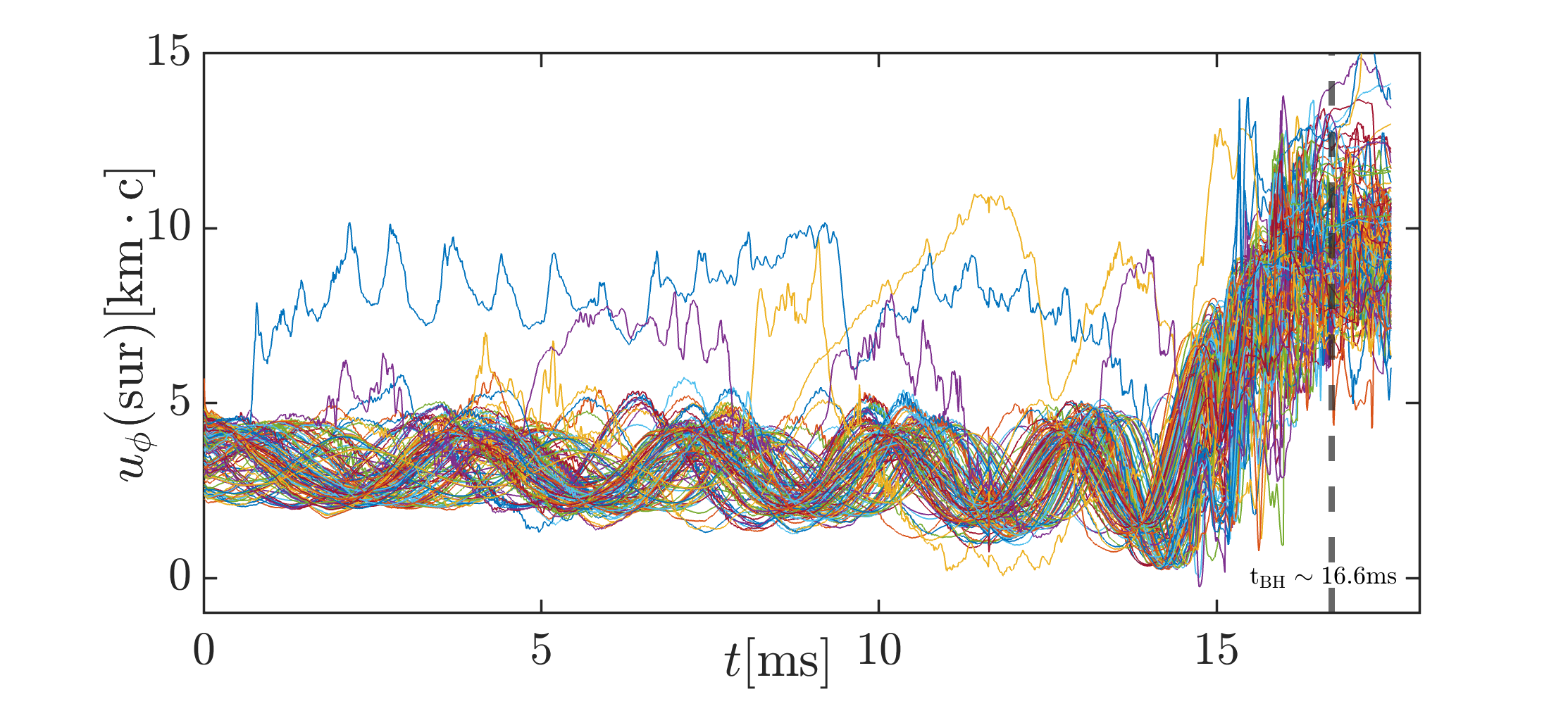}
\caption{Evolution of $u_{\phi}$, the conserved angular momentum component in an axisymmetric spacetime, for a large sample of the survivor tracer particles: $220$ particles selected at random. The small number of surviving particles with higher $u_{\phi}$ and noisy $u_\phi$ histories (blue and green lines) were initially located very close to the surface of one or the other of the NSs.}
\label{fig:uphi_sur_all}
\end{figure}

That these results are characteristic of \emph{all} the surviving particles are shown in Figure~\ref{fig:uphi_sur_all}. A small number of particles have large excursions in $u_\phi$ during the inspiral, but the overwhelming majority of particles follow the pattern of the small sample we have chosen for a particular study.

The torques creating the change in angular momentum could be the result of non-axisymmetric gravity \citep{Hotokezaka+13a, Radice+18_NSEoS, Fujibayashi+18, Shibata+23}, smooth pressure gradients, or shocks \citep{Hotokezaka+11PRD, Hotokezaka+13a, Sekiguchi+16}. If the magnetic field is as strong as \citet{Kiuchi+14} found (but not seen here), magnetic forces might also contribute.
Shocks can be recognized by the sharp steps in angular momentum evident in Figure~\ref{fig:uphi_compare} as well as the cusps seen in the trajectories of Figure~\ref{fig:XY_cm}, which might be associated with shock features visible in the later images of Figure~\ref{fig:Eint_xy_xz}. Both gravitational acceleration and smooth pressure gradients could lead to smooth portions of trajectories. The short timescale smaller fluctuations in $u_\phi$ shown in Figure~\ref{fig:uphi_compare} are more likely associated with pressure gradients than with gravity because there are many short length-scale density gradients, whereas the gravitational field is the result of the global stress-energy distribution. The general upward trend could be associated with non-axisymmetric gravity, and its end at $t\approx 16$~ms roughly coincides with the axisymmetrizing of the merged NS (see Fig.~\ref{fig:Ye_Rho_B_xy}). On the other hand, it is unclear how a gravitational effect would distinguish the particles that ultimately survive from the nearby particles that are ultimately captured; non-axisymmetric gravity could become important only after other dynamics place fluid elements in the regions where gravitational torques exist. In this context, it is also worth noting that the captured particles show much less short timescale fluctuation in $u_\phi$, suggesting that the regions through which they travel have much smoother density and pressure distributions. This would, of course, be a natural consequence of staying closer to the merged star.

\section{Discussion} \label{sec:discussion}

\subsection{Initial electron fraction in the bound debris}

The electron fraction $\ye$, the ratio of the number of protons $n_p$ to the total number of nucleons, is one of the key quantities for determining the elements synthesized over longer timescales in the bound debris. Through this influence, it is also the key to determining the opacity of the optical/UV photosphere where the KN emission is radiated \citep[see,][]{BarnesKasen13, Hotokezaka+13a, SiegelMetzger17PRL, Radice+18}.

If there were special locations in the NSs that contributed disproportionately to the bound debris, the distribution of mass with electron fraction $dM/d\ye$ in its initial state could be quite different from $dM/d\ye$ in the NSs as a whole. However, we have shown that, in rough terms, the probability that a fluid element inside one of the NSs is expelled from the remnant is very nearly constant for all radii between $\simeq 1$ and $\simeq 8$~km, a region containing the great majority of mass in the star.
There are only two exceptions to this constant probability. One is a spherical shell $\simeq 1$~km thick at the outside of the star, where the probability of escape rises to a few times greater than elsewhere. The other is a sphere of radius $\sim 1$~km at the center of the star, where the escape probability is $\lesssim 0.1\times$ the value in the star's bulk. Consequently, $dM/d\ye$ in the pre-merger NSs should be a very good predictor of $dM/d\ye$ in the bound debris immediately after it leaves the merged NS.

On the other hand, it should also be borne in mind that $\ye$ can be affected by events happening during the merger and while the bound debris travel from the merged NS to its place in the surrounding disk. Shock heating in the merger and during the subsequent evolution of the merger remnant can raise the matter temperature above $\rm 10 MeV$. In such a high-temperature environment, electron-positron pair creation can substantially raise the rate of the reaction $e^+ + n \rightarrow p^+ + \bar{\nu}_e$, thereby increasing $\ye$. Neutrino irradiation, within the merged star, as the gas travels to the disk, and in the disk can also raise $\ye$ via reactions like $\nu_e + n \rightarrow p^+ + e^-$. The degree to which any of these mechanisms may or may not alter $Y_e$ requires further investigation \citep{Shibata+23}.

\subsection{Parameter-dependence}

Quantitative details in the single calculation on which we base our results could change for different values of the principal system parameters, the mass ratio $q$ and the total mass $M$. In particular, the bilateral symmetry so evident in the images of Figures~\ref{fig:Ye_Rho_B_xy}, \ref{fig:Surv_nearst_dens}, and \ref{fig:XY_XZ_t1466_lx_Z} will be broken when $q \neq 1$. It has also long been known \citep{Oechslin_Janka06, Shibata+23} that the total mass removed from the merger remnant increases for smaller values of $q$, so the total amount of bound debris mass in this simulation is on the lower end of what might be expected for a merger of this total mass.

The results should also depend on $M$. If it is so large that the merger results in immediate collapse to a BH, the amount of expelled mass, both in dynamical ejecta and bound debris, would be much smaller. Moreover, because the tidal gravity contribution to the expelled mass would be a larger fraction of the total, the bound fraction would likely be considerably smaller. On the other hand, if $M$ were small enough for the merged NS to survive for a longer time, the expelled mass could be rather larger.

The combination of different $M$ and different $q$ can alter the results quantitatively in several ways. Smaller values of both imply particularly small mass for the less-massive star, giving it both lower mean density and a less centrally-concentrated internal density profile \citep[see,][]{LattimerPrakash07_EoS, Bauswein+12_EoS, Shibata+21, RaithelC+22_EoS}. Relatively large tidal mass-loss would likely follow. Larger $M$ and smaller $q$ could enhance the contribution of the contact shock if the larger mass star has $\gtrsim 2 M_\odot$ because for such large masses, the radius of the NS diminishes, whereas it hardly changes (for most EOSs) for masses between $\sim 0.7$ and $\sim 2M_\odot$ \citep{Kashyap+2022}. The sense in which these changes tip the balance between bound and unbound debris requires a detailed calculation to determine.

\subsection{Uncertainty in the equation of state}

Although the gravitational waveform measured in the event GW170817 has helped prune the long list of proposed nuclear EOSs \citep[see,][]{Hotokezaka+11PRD, Bauswein_Baumgarte_Janka13, Radice+18_NSEoS}, there still remain many candidates. Our quantitative results could be altered in several ways if a different EOS applies. First, the internal density profile of an equilibrium NS is sensitive to the EOS, and that could alter the probability of escape as a function of radius. Second, when the two stars come into contact, the pressure of the shocked region between the two depends on the EOS. Stiffer EOSs support larger NSs that strike each other with lower velocity, diminishing the post-shock temperature \citep{Hotokezaka+11PRD}; on the other hand, a larger ratio between pressure and internal energy density would make the shocked region wider and enlarge the pressure contrast between the shocked region and the surface of the merged star. Third, the EOS governs both the critical mass dividing quick collapse from a longer-lived merger NS and the time to collapse if rapid collapse is, indeed, the result \citep{Bauswein_Baumgarte_Janka13, Radice+18_NSEoS, Margalit+22}.
A longer lifetime for the merged NS likely leads to more escaping mass, and the ratio of bound to unbound mass could change over this span of time. Thus, different EOSs could lead to changes in both the total mass removed from the remnant and how it is divided into bound and unbound portions.

\section{Conclusions} \label{sec:conclusion}

In much of the literature on NS mergers, when matter that escapes being confined within the merger remnant is discussed, the focus is solely on the unbound fraction.   This fraction is then further divided into two categories, ``dynamical" and ``secular" (sometimes called ``post-merger") \citep{Metzger+10_KN, Fernandez+15, SiegelMetzger17PRL, Fujibayashi+18, Radice+18}. ``Dynamical" refers to the material expelled during the merger itself, whether by tidal forces, shocks heating gas, or neutrinos that heat or push the gas. ``Secular" refers to matter driven off the disk of bound debris many milliseconds after the merger. 

However, the majority of the matter outside the merger remnant remains bound to the system, forming the disk that is the source of the secular ejecta and may also support any jet that forms. Hitherto, the mechanisms by which this material leaves the merging neutron stars and is deposited in the debris disk have gone largely unexamined.
Here we have determined the origin of the bound debris within the NSs and examined how its fate is distinguished from that of matter remaining inside the merged star.

In particular, by making use of the complementary information provided by Eulerian fluid data and Lagrangian tracer particle data, we have uncovered several surprising facts about the process by which matter is expelled from merging neutron stars.

\begin{itemize}

\item  The probability that a given parcel of mass leaves the merger remnant is almost, but not quite, independent of its location inside the pre-merger NSs. Matter placed near the surface of a NS is considerably more likely than most matter to be expelled, but there is so much more matter in the bulk of the star that, despite its lower escape probability, nearly all the escaping matter, whether bound or unbound, has its origin deep in a star's interior.

\item Turbulent motions within the ``seam" surface formed when the two stars touch can mix matter from a wide range of initial depths within the stars. These same motions transport some of this matter through the seam plane to the surface of the merged star.

\item A significant amount of matter is brought out to the surface of the merged NS but does not escape. Which matter is captured and which escapes depends on details of the shock structure in the region where the matter that has been squeezed outward by the shock between the two merging stars reaches the surface.

\item The surviving matter acquires significant additional angular momentum as it leaves the remnant. It is likely that torques due to the non-axisymmetric gravity of the merged NS, smooth pressure gradients, and shocks all contribute. 

\item  All of these processes occur within a few milliseconds after the stars first touch.

\end{itemize}

Thus, we have shown that the process by which matter escapes the remnant created by a neutron star merger is considerably more complex and subtle than previously envisioned.
Because the disk of bound matter circulating around the merged NS or BH is far from inflow equilibrium, its subsequent evolution depends on what sort of non-equilibrium structure it is given. Understanding the mechanisms regulating this structure is, therefore, an essential part of understanding many of the dramatic observational properties of NS mergers---jets, kNe, and $r-$process nucleosynthesis.

\section*{Acknowledgements}
\begin{acknowledgments} \label{Sec:ack}
This work was partially supported by NASA TCAN grant NNH17ZDA001N. YZ thanks Prof. Will Farr, Prof. Yuri Levin, and Mor Rozner for their helpful discussions. L.R.W.\ and Z.B.E.\ gratefully acknowledge support from NSF awards PHY-1806596, PHY-2110352, OAC-2004311, as well as NASA award ISFM-80NSSC18K0538. A.M-B.\ is supported by NASA through the NASA Hubble Fellowship grant HST-HF2-51487.001-A awarded by the Space Telescope Science Institute, which is operated by the Association of Universities for Research in Astronomy, Inc., for NASA, under contract NAS5-26555.  
 T.P. was supported in part by ERC advanced grant ``Multijets".  This research made use of Idaho National Laboratory computing resources which are supported by the Office of Nuclear Energy of the U.S. Department of Energy and the Nuclear Science User Facilities under Contract No. DE-AC07-05ID14517.
\end{acknowledgments}

\software{astropy \citep{Astropy+18}, 
          \igm \citep{Etienne+15}, Numpy \citep{2020NumPy-Array}, and Matplotlib \citep{Hunter07_matplotlib}.}

\bibliography{refBNS23}{}
\bibliographystyle{aasjournal}



\end{document}